\documentclass[reprint, longbibliography, noeprint, superscriptaddress]{revtex4-1}

\usepackage[utf8]{inputenc}

\usepackage[lining,semibold]{libertine} 
\usepackage[libertine, cmintegrals, bigdelims, vvarbb]{newtxmath}

\usepackage{amsmath}
\usepackage{amsfonts}
\usepackage{mathrsfs}
\usepackage{gensymb}
\usepackage{bbm}
\usepackage{dsfont}

\usepackage{kbordermatrix}
\renewcommand{\kbldelim}{(}
\renewcommand{\kbrdelim}{)}

\usepackage{chemformula}
\usepackage[caption=false]{subfig}
\usepackage{chemfig}
\usepackage[version=3]{mhchem}

\usepackage{tikz}
\usetikzlibrary{matrix,positioning,decorations.pathreplacing}

\usepackage{soul}
\usepackage{xcolor}

\usepackage{scalerel} 
\usepackage{comment}

\usepackage{graphicx} 

\usepackage{bigints} 

\definecolor{webgreen}{rgb}{0,.5,0}
\definecolor{WEBGREEN}{rgb}{0,.5,0}
\definecolor{webbrown}{rgb}{.6,0,0}
\definecolor{WEBBROWN}{rgb}{.6,0,0}
\definecolor{grigio}{rgb}{.85,.85,.85} 
\definecolor{RoyalBlue}{rgb}{0.0, 0.14, 0.4}
\definecolor{skyblue1}{rgb}{0.45,0.62,0.81}
\definecolor{skyblue2}{rgb}{0.2,0.39,0.64}
\definecolor{skyblue3}{rgb}{0.13,0.29,0.53}
\definecolor{scarlet1}{rgb}{0.93,0.16,0.16}
\definecolor{SCARLET1}{rgb}{0.93,0.16,0.16}
\definecolor{scarlet2}{rgb}{0.8,0,0}
\definecolor{scarlet3}{rgb}{0.64,0,0}

\definecolor{g}{gray}{0.50}

\usepackage{hyperref}
\hypersetup{%
    colorlinks=true, linktocpage=true, pdfstartpage=1, pdfstartview=FitV,%
    breaklinks=true, pdfpagemode=UseNone, pageanchor=true, pdfpagemode=UseOutlines,%
    plainpages=false, bookmarksnumbered, bookmarksopen=true, bookmarksopenlevel=1,%
    hypertexnames=true, pdfhighlight=/O,
    urlcolor=webbrown, linkcolor=RoyalBlue, citecolor=webgreen, 
    pdftitle={},%
    pdfauthor={Francesco Avanzini},%
    pdfsubject={},%
    pdfkeywords={},%
    pdfcreator={pdfLaTeX},%
    pdfproducer={LaTeX REVTeX}%
}

\newcommand{\remark}{\textit{Remark. }}
\newcommand{\notation}{\textit{Notation. }}

\newcommand{\rev}[1]{{\color{black}#1}}

\newcommand{\fref}[1]{{\begingroup\footnotesize(\ref{#1})\endgroup}}

\newcommand{\transition}{transition}
\newcommand{\transitions}{transitions}
\newcommand{\Transition}{Transition}
\newcommand{\Transitions}{Transitions}

\newcommand{\reaction}{reaction}
\newcommand{\reactions}{reactions}
\newcommand{\Reaction}{Reaction}
\newcommand{\Reactions}{Reactions}

\newcommand{\isomerization}{isomerization}

\newcommand{\Isomerization}{Isomerization}

\newcommand{\futile}{futile}
\newcommand{\Futile}{Futile}

\newcommand{\ratchet}{{ratchet}}
\newcommand{\ratchets}{{ratchets}}

\newcommand{\Ratchets}{{Ratchets}}

\newcommand{\process}{{process}}
\newcommand{\processes}{{processes}}
\newcommand{\Process}{{Process}}

\newcommand{\steady}[1]{{#1}_{\text{ss}}}
\newcommand{\equi}{{\text{eq}}}

\newcommand{\dt}{\mathrm d_t}
\newcommand{\dd}{\mathrm d}

\newcommand{\ddirac}[2]{\delta({#1 - #2})}
\newcommand{\ddiracc}[1]{\delta({#1})}

\newcommand{\kb}{k_{\mathrm{B}}}

\newcommand{\gs}{\alpha}
\newcommand{\gss}{\beta}

\newcommand{\ggs}{\alpha'}
\newcommand{\ggss}{\beta'}

\newcommand{\es}{i_{\gs}}
\newcommand{\ess}{j_{\gss}}

\newcommand{\egs}[1]{i_{#1}}
\newcommand{\egss}[1]{j_{#1}}

\newcommand{\Ze}{\ch{X}^{}_{\es}}
\newcommand{\Zee}{\ch{X}^{}_{\ess}}

\newcommand{\Zg}{\ch{X}^{}_{\gs}}
\newcommand{\Zgg}{\ch{X}^{}_{\gss}}

\newcommand{\Zs}{\ch{X}_{\mathrm{solv}}}

\newcommand{\pp}{\mathrm{p}}
\newcommand{\p}{\pp_\nu}

\newcommand{\fe}{\nu_{\es, \ess}}
\newcommand{\fevib}{\nu_{\egss{\gs}, \egs{\gs}}}
\newcommand{\pe}{\pp_{\fe}}

\newcommand{\n}[1]{n(#1)}
\newcommand{\denp}[1]{\varrho(#1)}

\newcommand{\Tbb}[1]{T^{\mathrm{bb}}(#1)}
\newcommand{\TTbb}{T^{\mathrm{bb}}}

\newcommand{\kq}[2]{k^{\mathrm{q}}_{#1\to#2}}

\newcommand{\qflux}[2]{R^{\mathrm{q}} ({#1\to#2})}

\newcommand{\qcurr}[2]{J^{\mathrm{q}} ({#1\to#2})}

\newcommand{\kcd}[3]{k^{\mathrm{#3}}_{#1\to#2}}

\newcommand{\cdflux}[3]{R^{\mathrm{#3}} ({#1\to#2})}

\newcommand{\kqeq}[2]{K^{\mathrm{\equi}}_{#1\to#2}}

\newcommand{\qcflux}[3]{\Pi^{\mathrm{q}} \Big({#1\xrightarrow{#3}#2}\Big)}

\newcommand{\qccurr}[3]{\Psi^{\mathrm{q}} \Big({#1\xrightarrow{#3}#2}\Big)}

\newcommand{\kpa}[2]{k^{\mathrm{a}}_{#1\to#2}}
\newcommand{\kpe}[2]{k^{\mathrm{e}}_{#1\to#2}}
\newcommand{\kpse}[2]{k^{\mathrm{se}}_{#1\to#2}}

\newcommand{\dkpa}[3]{k^{\mathrm{a}}_{#1\to#2}(#3)}
\newcommand{\dkpe}[3]{k^{\mathrm{e}}_{#1\to#2}(#3)}
\newcommand{\dkpse}[3]{k^{\mathrm{se}}_{#1\to#2}(#3)}

\newcommand{\pfluxa}[2]{R^{\mathrm{a}}({#1\to#2})}
\newcommand{\pfluxe}[2]{R^{\mathrm{e}}({#1\to#2})}

\newcommand{\pdfluxa}[3]{r^{\mathrm{a}}({#1\to#2}, #3)}
\newcommand{\pdfluxe}[3]{r^{\mathrm{e}}({#1\to#2}, #3)}

\newcommand{\pcurr}[2]{J^{\mathrm{p}}({#1\to#2})}

\newcommand{\pdcurr}[3]{j^{\mathrm{p}}({#1\to#2, #3})}

\newcommand{\excurr}{I(\nu)}

\newcommand{\pcdfluxa}[4]{\pi^{\mathrm{a}}\Big({#1\xrightarrow{#3}#2}, #4\Big)}
\newcommand{\pcdfluxe}[4]{\pi^{\mathrm{e}}\Big({#1\xrightarrow{#3}#2}, #4\Big)}

\newcommand{\pcddflux}[5]{\pi \Big({#1\xrightarrow{#3}#2}, #4, #5\Big)}

\newcommand{\pcdcurr}[4]{\psi^{\mathrm{p}}\Big({#1\xrightarrow{#3}#2, #4}\Big)}
\newcommand{\pcddcurr}[5]{\psi^{\mathrm{p}}\Big({#1\xrightarrow{#3}#2, #4, #5}\Big)}

\newcommand{\conc}[1]{[#1]}
\newcommand{\ssconc}[1]{\steady{[#1]}}

\newcommand{\npt}[2]{n(\nu_{#1,#2})}
\newcommand{\np}{n(\nu)}

\newcommand{\prob}[2]{p(#1|#2)}

\newcommand{\partition}[1]{Q_{#1}}

\newcommand{\stie}[1]{u^\circ_{#1}}

\newcommand{\sts}[1]{s^\circ_{#1}}

\newcommand{\cp}[1]{\mu^{}_{#1}}
\newcommand{\stcp}[1]{\mu^\circ_{#1}}
\newcommand{\sttcp}[1]{\hat{\mu}^\circ_{#1}}

\newcommand{\cpp}[1]{\mu^{}_{\mathrm{p}}(#1)}
\newcommand{\cppbb}{\mu^{\mathrm{bb}}_{\mathrm{p}}}

\newcommand{\free}{F}
\newcommand{\freeeq}{F_{\mathrm{eq}}}

\newcommand{\freec}{F_{\mathrm{ch}}}
\newcommand{\freep}{F_{\mathrm{ph}}}
\newcommand{\dfreep}{f_{\mathrm{ph}}(\nu)}

\newcommand{\epr}{\dot{\Sigma}}

\newcommand{\eprq}{\dot{\Sigma}_\mathrm{q}}
\newcommand{\eprp}{\dot{\Sigma}_\mathrm{p}}

\newcommand{\deprp}{\dot{\sigma}_\mathrm{p}(\nu)}

\newcommand{\pw}{\dot{W}_\mathrm{p}}

\newcommand{\DD}[1]{\mathcal{D}_{{#1}}}

\newcommand{\qyq}[2]{\Phi^{\mathrm{q}}(#1\to#2)}
\newcommand{\qyp}[3]{\phi^{\mathrm{p}}(#1\to#2,#3)}

\newcommand{\cZ}{\ch{Z}}
\newcommand{\cZZ}{\ch{Z^{*}}}
\newcommand{\cE}{\ch{E}}
\newcommand{\cEE}{\ch{E^{*}}}

\newcommand{\cZa}{\ch{Z_1}}
\newcommand{\cZb}{\ch{Z_2}}
\newcommand{\cZZa}{\ch{Z^{*}_1}}
\newcommand{\cZZb}{\ch{Z^{*}_2}}

\newcommand{\cEa}{\ch{E_1}}
\newcommand{\cEb}{\ch{E_2}}
\newcommand{\cEEa}{\ch{E^{*}_1}}
\newcommand{\cEEb}{\ch{E^{*}_2}}

\newcommand{\cF}{\ch{F}}
\newcommand{\cW}{\ch{W}}

\newcommand{\Kr}{K_r}

\makeatletter
\def\maketag@@@#1{\hbox{\m@th\normalfont\normalsize#1}}
\makeatother

\DeclareMathAlphabet{\mathpzc}{OT1}{pzc}{m}{it}

\begin{document}

\title{Coarse Graining Photo-Isomerization Reactions: \\
Thermodynamic Consistency and Implications for Molecular \Ratchets
}

\newcommand\unipdchem{\affiliation{Department of Chemical Sciences, 
University of Padova, 
Via F. Marzolo, 1, I-35131 Padova, Italy}}
\newcommand\unilu{\affiliation{Complex Systems and Statistical Mechanics, 
Department of Physics and Materials Science, 
University of Luxembourg, 
30 Avenue des Hauts-Fourneaux, L-4362 Esch-sur-Alzette, Luxembourg}}
\newcommand\unipdphy{\affiliation{Department of Physics and Astronomy ``Galileo Galilei'',
University of Padova, 
Via F. Marzolo, 8, I-35131 Padova, Italy}}
\newcommand\infn{\affiliation{INFN, Sezione di Padova, 
via Marzolo 8, I-35131 Padova, Italy}}
\newcommand\uninw{\affiliation{Department of Chemistry, 
Northwestern University, Evanston, IL 60208, USA}}

\author{Francesco Avanzini}
\email{francesco.avanzini@unipd.it}
\unipdchem
\author{Massimiliano Esposito}
\email{massimiliano.esposito@uni.lu}
\unilu
\author{Emanuele Penocchio}
\email{emanuele.penocchio@northwestern.edu}
\uninw


\date{\today}

\begin{abstract}
We formulate thermodynamically consistent coarse-graining procedures 
for molecular systems
undergoing thermally and photo-induced \transitions:
starting from elementary vibronic \transitions, 
we derive effective photo-\isomerization\ \reactions\ 
interconverting ground-state species.
Crucially, 
the local detailed balance condition, 
that constrains reaction kinetics to thermodynamics, 
remains satisfied throughout the coarse-graining procedures.
It applies to the effective photo-isomerization reactions just as it does to the elementary vibronic transitions.
We then demonstrate that 
autonomous
photo-driven molecular \ratchets\ 
operate via 
the same fundamental mechanism 
as chemically driven ones.
Because the local detailed balance remains satisfied,
autonomous
photo-driven molecular \ratchets, like chemically driven ones,
operate exclusively through an information ratchet mechanism.
\rev{This reveals that 
their design and optimization should prioritize 
molecular properties governing the information ratchet mechanism, 
rather than those influencing energetic bias.}
\end{abstract}

\maketitle


\section{Introduction}

Molecular systems are dissipative systems 
whose dynamics is ultimately governed by elementary transitions,
interconverting  molecular states and species,
that must obey thermodynamic constraints.
The fundamental constraint that all elementary transitions must satisfy
is the local detailed balance condition~\cite{lebowitz1955,maes2021,Peliti2021}:
the fluxes of each pair of forward and backward transitions do not take arbitrary values
but are constrained by the corresponding thermodynamic force, i.e., the free energy variation along the transition.
This condition guarantees that molecular systems relax to equilibrium
unless they continuously harness free energy 
from their surroundings by exchanging energy and/or chemical species~\cite{postmodern}.
It also ensures that 
the contribution of all elementary transitions to the total dissipation
can be properly accounted for.

Molecular systems that are maintained out of equilibrium 
through the exchanges of chemical species with their surroundings
are said to be chemically driven.
In this context,
the local detailed balance
enabled the development of general nonequilibrium thermodynamic theories~\cite{Gaspard2004, Schmiedl2007, Rao2018b, Qian2005, Rao2016, Avanzini2021, avanzini2022}.
These theories can quantify the energetic cost of, for instance,
maintaining coherent oscillations~\cite{Oberreiter2022b, Remlein2022, Ohga2023, santolin2025} 
and patterns~\cite{Falasco2018a, Avanzini2019a, Avanzini2024rds, voss2025chemomechanicalmotilitymodespartially},
as well as powering chemical growth~\cite{Marehalli2024prl, Marehalli2024pre}.
They can also identify
the accessible chemical space~\cite{LiangBusiello2024, liang2024chemspace},
the information encoded in polymers~\cite{Andrieux2008, Blokhuis2017},
and speed limits~\cite{Yoshimura2021a, Yoshimura2021b}.
Furthermore, 
they have been used to characterize the energetics of specific systems,
like
the efficiency of molecular motors~\cite{amano2021info,Penocchio2022,sivak2025revflow}
and the energy stored by driven endergonic processes~\cite{penocchio2019eff,prins2023energystored,leigh2024endosynth,ragazzon2024endosynth,ragazzon2025activetransport}.
However, 
their direct application becomes quickly impractical 
when molecular systems reach high levels of complexity 
arising from a large number of species interconverted by a vast network of transitions.
Indeed, they require knowing 
the concentrations of all species and the rate constants of every elementary transition,
which is very seldom achieved.
To address this issue, 
\rev{different strategies have been developed that either infer, from incomplete experimental data, 
the structure of the underlying network of transitions~\cite{Blokhuis2025, blokhuis2025datadimensionchemistry} 
and the associated thermodynamic quantities~\cite{Harunari2022, vanderMeer2022,Maier2024},
or provide thermodynamically consistent coarse-graining procedures~\cite{Wachtel2018, Avanzini2020b, Avanzini2023, Raux2024}.
In particular, these  coarse-graining procedures yield simplified dynamic descriptions of complex molecular systems} 
while retaining the exact dissipation by identifying effective transitions between slow-evolving species.
Notably, thermodynamic consistency is retained 
even if the fluxes of the effective transitions do not, in general, satisfy the local detailed balance condition 
hence hiding this fundamental constraint.
These coarse graining procedures constitute a powerful tool
to characterize the energetics of systems regardless of their size.
Indeed, 
they have been used to study
the efficiency of processes at the cellular scale~\cite{Wachtel2022, Voorsluijs2024},
but some recent attempts also explored applications at the ecosystem scale~\cite{Goyal2023}.

Molecular systems that are instead maintained out of equilibrium through the exchanges of energy with light
are said to be photo-driven.
In this case, 
the photo-induced elementary transitions 
are intramolecular vibronic transitions, 
i.e., absorption, spontaneous, and stimulated emission of photons
interconverting vibrational and electronic states of a molecule~\cite{Balzani2014}.
The local detailed balance condition still holds
and
corresponds to Einstein's relations~\cite{Cohen-Tannoudji1998},
which laid the foundation for a thermodynamic understanding of 
light-matter interactions.
Nevertheless,
thermodynamic theories for photo-driven molecular systems are not as well established
as those for chemically driven ones.
The current formulation considers only two-state molecules that interact with monochromatic light~\cite{penocchio2021photo}.
Molecular species are instead multi-state systems 
whose elementary photo-induced dynamics is driven by polychromatic light 
(with a continuous spectrum).
Furthermore, many fast-evolving excited-state species, 
which are challenging to characterize experimentally,
are involved in photo-induced dynamics,
thus underscoring the need for thermodynamically consistent coarse-graining theories.

In this paper, we develop such thermodynamic consistent theories for photo-driven molecular systems.
We begin by establishing a general framework for elementary transitions interconverting 
molecular species with an arbitrary number of vibrational and electronic states 
interacting with polychromatic light~(Sec.~\ref{sec:elementary}).
We then develop two sequential coarse-graining procedures.
The first one coarse grains the vibrational states~(Sec.~\ref{sec:cg_vibrational_states}), 
yielding a description in terms of electronic states only.
The second one coarse grains electronic excited states~(Sec.~\ref{sec:HDM}),
yielding a description in terms of ground-state species only.
This enables us to define effective photo-isomerization reactions~(Sec.~\ref{sec:DM})
for a paradigmatic photochemical process, 
namely, the diabatic process~\cite{foerster1970}.
Crucially, 
both these coarse-graining procedures are thermodynamically consistent, 
preserving the exact dissipation, 
and furthermore still satisfy the local detailed balance condition.
On the one hand,
thermodynamic consistency (Subs.~\ref{sub:SL1} and~\ref{sub:SL2}) is achieved by 
defining the (effective) photo-isomerization reactions 
in terms of cycles between the ground-state species, 
similarly to what has been done for the coarse-graining of chemically driven systems~\cite{Wachtel2018,Avanzini2020b}.
On the other hand,
the local detailed balance (Subs.~\ref{sub:LDB1} and~\ref{sub:LDB2}) is retained by 
using the cycles identified by a modified version of the matrix-tree theorem~\cite{Hill1966, HILL197733} 
already used in Ref.~\cite{LiangBusiello2024}.
This is a major difference from the coarse-graining of chemically driven systems 
as it implies that the fluxes of the (effective) photo-isomerization reactions,
not just those of the underlying elementary \transitions,
are also constrained by the corresponding free energy variation.
The result is that both the dynamics and the energetics of photo-driven systems 
can be characterized in terms of experimentally accessible quantities: 
the former through absorption spectra and quantum yields,
while the latter through the chemical potentials of the photons and of the ground-state species.

The importance of having such thermodynamic theories cannot be overstated,
given the ubiquity of photo-driven systems
in inorganic~\cite{Ghil2020} and organic matter~\cite{Blankenship2014}
\rev{as well as the increasing attention they are receiving in current research.}
In biology, 
prominent examples of these systems include 
photosynthesis~\cite{Blankenship2014}, transducing photons into chemical energy, 
and animal vision~\cite{cerullo2010}, where photons with certain frequencies initiate a signaling cascade.
In systems chemistry,
prominent examples include 
several classes of molecular motors~\cite{feringa1999,pooler2021}, pumps~\cite{leigh2007,Corra2022a}, and active materials~\cite{giuseppone2015,giuseppone2017}, transducing photons into 
nanoscale directional motion or even macroscale work,
as well as enhancement of catalytic activity under nonequilibrium conditions~\cite{prins2020catalysis}, and active transport through a liquid membrane~\cite{nitschke2024activetransport,aprahamian2024activetransport}.

All these nonequilibrium chemical systems 
epitomize a wide class of systems known as 
molecular ratchets~\cite{Ragazzon2023review,Borsley2024,astumian2024review}.
\rev{While no universally accepted definition exists
(e.g., 
the ones in Refs.~\cite{Ragazzon2023review} and~\cite{Borsley2024} overlap, 
but they are not equivalent),
we adopt here the following definition.
Molecular ratchets are systems that}
couple 
free-energy-harnessing transitions 
to endergonic ones
in a way 
that \rev{drives the latter against their spontaneous direction
by exploiting an energy source 
(e.g., chemical potential gradients or light) 
with which only the former are in direct contact~\cite{Ragazzon2023review}.}
It is currently believed
that 
photo-driven molecular ratchets operate via a different mechanism 
with respect to chemically driven ones~\cite{astumian2016,astumian2019,Ragazzon2023review,Borsley2024,astumian2024review, corra2025review}.
In this paper, 
we show that this is not the case (Sec.~\ref{sec:MM})
by applying our theories, and
in particular the local detailed balance condition 
of the (effective) photo-isomerization reactions, 
to photo-driven molecular ratchets.
Namely, chemically and photo-driven ratchets 
operate via a pure information ratchet mechanism when powered by a constant energy source 
(named autonomous ratchets), 
thus unifying their description 
and unraveling a paradigmatic change in the design and optimization principles 
of photo-driven ratchets.

We conclude by 
briefly contextualizing our work 
within the historical development of the local detailed balance condition 
in Sec.~\ref{sec:micro_rev} 
and 
by summarizing and further discussing our results 
in Sec.~\ref{sec:conclusions}.


\section{Elementary \Transitions \label{sec:elementary}}

\begin{figure}[t]
    \centering
    \includegraphics[width=0.49\textwidth]{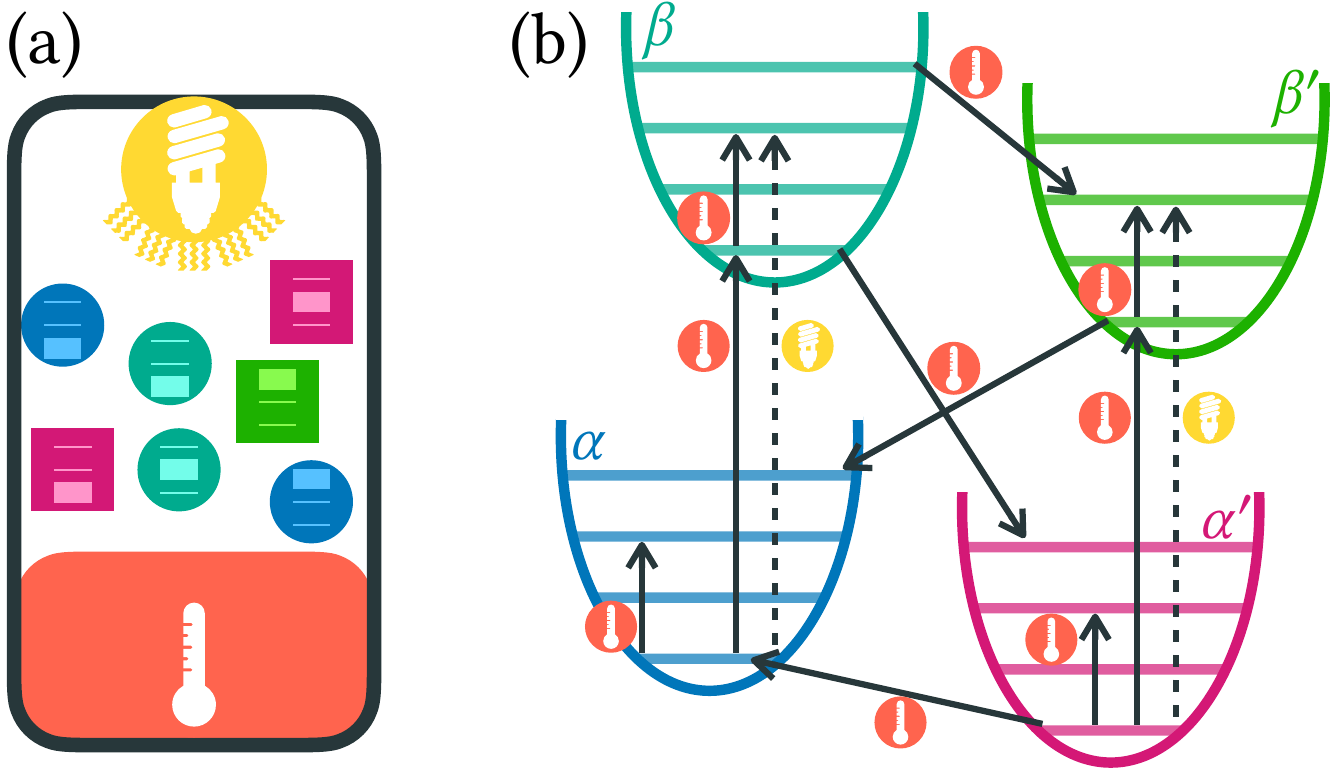}
    \caption{
    (a)
    Pictorial illustration of an ideal dilute solution 
    maintained at constant temperature by the solvent (represented by a thermometer)
    and immersed in incoherent
    light generated by a light source (represented by a bulb).
    Two different molecular species (represented by disks and squares) can exist in different 
    electronic (represented by different colors) and 
    vibrational (represented by the different position of the thickest line) states.
    (b) 
    Some \transitions\ (arrows) between different vibrational and electronic states of two species. 
    The first (resp. second) species exists in two electronic states $\gs$ and $\gss$ (resp. $\ggs$ and $\ggss$).
    \Transitions\ interconverting states of the same species (vertical arrows) can be 
    thermally induced (arrows close to a thermometer) 
    and/or photo-induced (dashed arrows close to a bulb).
    \Transitions\ interconverting states of different species (oblique arrows) can only be 
    thermally induced (arrows close to a thermometer).
    Each \transition\ has always a backward counterpart, even if it is not represented.
    \label{fig:basic}}
\end{figure}

We consider a homogeneous ideal dilute solution with volume $V$
immersed in incoherent light (photons) as illustrated in Fig.~\ref{fig:basic}.
The solution is 
maintained at constant temperature $T$ 
by the solvent playing the role of a thermostat,
while photons are
generated by a light source.
Different molecular species exist in different vibrational and electronic states,
while photons can have different frequencies.
Vibronic \transitions\ within the same species and from one species to another 
can be thermally induced,
namely, they can be promoted by the interactions with the thermostat.
Vibronic \transitions\ within the same species 
can be photo-induced too,
namely, they can be promoted by the absoption/emission of photons.

\notation
In the following, 
greek letters $\gs$ or $\gss$ specify 
both the molecular species and its electronic state. 
In the rare case where we need to explicitly distinguish the electronic states of two different species, 
$\ggs$ or $\ggss$ are used for the second one.
On the other hand, 
$\egs{\gs}$ or $\egss{\gs}$ (resp.  $\egs{\gss}$ or $\egss{\gss}$) specify
the vibrational state corresponding to the species and electronic state $\gs$ (resp. $\gss$).
The chemical symbol and concentration of a species in a specific vibrational state are given by 
$\Ze$ and $\conc{\Ze}$,
respectively.
The chemical symbol and concentration of a species in a specific electronic state,
regardless of the vibrational state,
are given by 
$\Zg$ and $\conc{\Zg} = \sum^{ }_{\es} \conc{\Ze}$,
respectively.
Photons with frequencies $\nu$ are represented by the symbol $\p$.
The concentration of photons with frequencies in the interval $[\nu,\nu+\dd\nu]$ is given by $\n{\nu}\dd\nu$,
while we will refer to $\n{\nu}$ as the density of photons with frequency~$\nu$.

\subsection{Kinetics\label{sub:el_kin}}

Each thermally induced \transition\ from $\es$ to $\ess$ 
has always its backward counterpart from $\ess$ to $\es$.
Together, these two \transitions\ can be represented by the chemical equations
\begin{subequations}
\begin{align}
{\Ze} &\ch{->} {\Zee}\,,\\
{\Zee} &\ch{->} {\Ze}\,,
\end{align}
\label{eq:qrct}
\end{subequations}
whose fluxes read
\begin{subequations}
\begin{align}
\qflux{\es}{\ess} &= \kq{\es}{\ess}\conc{\Ze}\,,\\
\qflux{\ess}{\es} &= \kq{\ess}{\es}\conc{\Zee}\,,
\end{align}
\label{eq:flux_q}
\end{subequations}
respectively, 
with $\kq{\es}{\ess}$ and $\kq{\ess}{\es}$ being the kinetic constants.
The corresponding net current is given by 
\begin{equation}
\qcurr{\es}{\ess} = \qflux{\es}{\ess} - \qflux{\ess}{\es}\,.
\label{eq:curr_q}
\end{equation}

\remark 
Thermally induced \transitions\ can interconvert any pair of vibrational states.
On the one hand,
they are usually referred to as quenching \transitions\ 
when they interconvert vibrational states of the same species, e.g., $\es$ and $\egss{\gss}$.
On the other hand,
they define (unimolecular) isomerization \transitions\
when they interconvert vibrational states of different species, e.g., $\es$ and $\egss{\ggs}$.

Each photo-induced \transition\ from $\es$ to $\ess$ of the same species
promoted by the absorption of a photon of frequency $\fe$ 
has always its backward counterpart in the spontaneous emission and stimulated emission from $\ess$ to $\es$~\footnote{
\rev{Note that each thermally induced \transition\ has its backward counterpart 
in the \transition\ corresponding to its flipped chemical equation. 
By contrast, a photo-induced \transition\ due to photon absorption has its backward counterpart in two \transitions : 
spontaneous or stimulated emission. 
Only spontaneous emission corresponds to the flipped chemical equation of absorption, 
whereas stimulated emission does not. 
This difference between thermally and photo-induced transitions is not a contradiction: 
in quantum electrodynamics, both spontaneous and stimulated emission originate from the same elementary interaction 
between matter and the quantized electromagnetic field 
and thus reflect two aspects of the same fundamental phenomenon}}.
Together, these three \transitions\ can be represented by the following chemical equations
\begin{subequations}
\begin{align}
\Ze + \pe &\ch{->} \Zee\,, \\
\Zee &\ch{->} \Ze + \pe\,, \\
\Zee + \pe &\ch{->} \Ze + 2\pe\,,
\end{align}
\label{eq:prct}
\end{subequations}
respectively.
Here, we assumed, without loss of generality, 
that the energy of state~$\ess$ is larger than the energy of state~$\es$,
i.e., $\stie{\ess} > \stie{\es}$.
Furthermore, the frequency $\fe$ corresponds exactly to the energy difference between the two states,
i.e., $h \fe = |\stie{\ess} - \stie{\es}|$ (with $h$ being the Planck constant).
The fluxes of absorption and emission read
\begin{subequations}
\begin{align}
\pfluxa{\es}{\ess} &= \kpa{\es}{\ess}\, \npt{\es}{\ess}\conc{\Ze}\,,\\
\pfluxe{\ess}{\es} &= \big(\kpe{\ess}{\es} + \kpse{\ess}{\es} \,\npt{\es}{\ess}\big)\conc{\Zee}\,,
\end{align}
\label{eq:flux_p}
\end{subequations}
where $\kpa{\es}{\ess}$, $\kpe{\ess}{\es}$, and $\kpse{\ess}{\es}$ are the Einstein's coefficients,
namely, the kinetic constants
for absorption, spontaneous emission, and stimulated emission, respectively~\cite{Cohen-Tannoudji1998}.
The corresponding net current  is given by
\begin{equation}
\pcurr{\es}{\ess} 
= \pfluxa{\es}{\ess} - \pfluxe{\ess}{\es}\,.
\label{eq:ecurrp}
\end{equation}
If $\stie{\ess} < \stie{\es}$, then $\pcurr{\es}{\ess} = \pfluxe{\es}{\ess} - \pfluxa{\ess}{\es}$.

\remark 
Photo-induced \transitions\ cannot interconvert states of different species, e.g., $\es$ and $\egss{\ggs}$,
as a result of 
the Franck-Condon principle~\cite{Balzani2014}
implying that $\pfluxa{\es}{\egss{\ggs}} = \pfluxe{\egss{\ggs}}{\es} = 0$.


\subsection{Rate Equations\label{sub:req}}

The concentration of a species in a specific state, i.e.,~$ \conc{\Zee}$, 
is affected by both thermally and photo-induced \transitions\
and thus evolves in time according to the following rate equation:
\begin{equation}
\dt \conc{\Zee} =
{\sum_{\gs}}
{\sum_{\substack{\es \\ (\es\neq\ess)}}}
\Big(
\qcurr{\es}{\ess} + \pcurr{\es}{\ess} 
\Big)\,.
 \label{eq:req}
\end{equation}

The density of photons is instead not affected by the thermally induced \transitions .
It is affected 
by the photo-induced \transitions\ and 
by the light source, the latter represented by the net density current $\excurr$
quantifying the net density of photon of frequency $\nu$ 
entering ($\excurr>0$) or exiting ($\excurr<0$) the solution.
Hence, the density of photons $\n{\nu}$ follows the rate equation~\footnote{
The rate equation~\fref{eq:req_p} is different
from the rate equation for the density of photons $\n{\nu}$ used in Ref.~\cite{penocchio2021photo}:
the Dirac delta function is missing in the latter.
This creates a missmatch of units between the left and the right hand side of the rate equation in Ref.~\cite{penocchio2021photo}
that does not anyway compromise the paper results since monochromatic light was considered.
}
\begin{equation}
\dt\n{\nu} = 
-
\sum_{\gs,\gss}
\sum_{\ess>\es} 
\ddirac{\nu}{\fe} \,
\pcurr{\es}{\ess} +  \excurr \,,
\label{eq:req_p}
\end{equation}
where $\ddiracc{\bullet}$ is the Dirac delta function
and we ordered the states, without loss of generality,
in such a way that $\ess > \es$ if $\stie{\ess} > \stie{\es}$.

\remark
In general, the net density current $\excurr$ is non-homogeneous in the solution.
In typical photochemical experiments
the solution is contained in a cuvette and irradiated from one side with a fix net density current~\cite{Montalti2006}.
According to the Beer-Lambert law, 
the magnitude of net density current then decreases along the direction of propagation
because photons are absorbed. 
Nevertheless, 
the generalization of our framework to a non-homogeneous net density current $\excurr$ is straightforward:
the solution can be split in small volumes where $\excurr$ is approximately homogeneous 
as already shown in Refs.~\cite{Penocchio2022, Corra2022a} for the case of monochromatic light.


\subsection{Local Detailed Balance\label{sub:ldb}}

The fundamental constraint that all elementary transitions must satisfy
is the local detailed balance condition~\cite{maes2021,Peliti2021,postmodern},
establishing a correspondence between dynamics and thermodynamics.

For each thermally induced \transition\,
the local detailed balance condition reads~\cite{Rao2016, postmodern}
\begin{equation}
\kb T\ln\frac{\qflux{\es}{\ess}}{\qflux{\ess}{\es}} = - \big( \cp{\ess} - \cp{\es}  \big) \,,
\label{eq:ldb_qrct} 
\end{equation}
where 
$\cp{\es}$ is the chemical potential~\footnote{
\rev{
The expression of the chemical potential in Eq.~\fref{eq:cp} is mathematically correct 
even if it features the logarithm of a dimensional quantity such as the concentration~$\conc{\Ze}$.
Indeed, the standard chemical potential $\stcp{\es}$ has been (implicitly) rescaled by the concentration of the solvent $\conc{\Zs}$
according to $\stcp{\es} = \sttcp{\es} - \kb T \ln \conc{\Zs}$ and, therefore,
the chemical potential reads $\cp{\es} = \sttcp{\es} + \kb T\ln\conc{\Ze}/\conc{\Zs}$ where the logarithm is a function of a dimensionless quantity as shown in Ref.~\cite{Fermi1956}}} 
of the species in state $\es$
\begin{equation}
\cp{\es} = \stcp{\es} + \kb T\ln\conc{\Ze}
\label{eq:cp}
\end{equation}
with $\stcp{\es} =\stie{\es} - T \sts{\es}$ the corresponding standard chemical potential 
and $\sts{\es}$ the corresponding standard entropy.
The local detailed balance condition~\eqref{eq:ldb_qrct} is the direct implication 
of the fact that thermally induced \transitions\ 
must admit an equilibrium state
when no chemical species are exchanged with the surroundings~\cite{tolman1925,lewis1925microrev,postmodern}.


For each photo-induced \transition\,
the local detailed balance can be written as~\cite{penocchio2021photo}
\begin{equation}
\kb T\ln \frac{\pfluxa{\es}{\ess}}{\pfluxe{\ess}{\es}} =
- \big( \cp{\ess} - \cp{\es} - \cpp{\fe} \big)\,,
\label{eq:ldb_prct}
\end{equation}
where we assumed, without loss of generality, that $\stie{\ess} > \stie{\es}$
and the chemical potential of the photons with frequency~$\nu$ reads
\begin{equation}
\cpp{\nu} = h \nu -\kb T \ln\frac{\denp{\nu}+\n{\nu}}{\n{\nu}}\,,
\label{eq:cpp}
\end{equation}
with $\denp{\nu}$ being the density of photon states associated to the radiation with frequency $\nu$
(reading $\denp{\nu} = 8\pi \nu^2 / c^3$ in vacuum,
with $c$ the speed of light)~\footnote{
\rev{
Note that evaluating the chemical potential of the photons with frequency~$\nu$ requires 
no detailed knowledge of the light source. 
It only requires knowledge of the density of photon states $\denp{\nu}$ and the density of photons $\n{\nu}$.}}.
The local detailed balance condition~\eqref{eq:ldb_prct} is a direct implication of the Einstein's relations~\cite{Cohen-Tannoudji1998,Hilborn1982}
\begin{subequations}
\begin{align}
\frac{\kpa{\es}{\ess}}{\kpse{\ess}{\es}}& = \exp\frac{\sts{\ess} - \sts{\es} }{\kb}\,,\label{eq:einstein1}\\
\frac{\kpe{\ess}{\es}}{\kpse{\ess}{\es}}& = \denp{\fe}\,,\label{eq:einstein2}
\end{align}\label{eq:einstein}%
\end{subequations}
as it can be proven by direct substitution of Eqs.~\eqref{eq:einstein},~\eqref{eq:cpp}, and~\eqref{eq:cp} into Eq.~\eqref{eq:ldb_prct}.

\remark
Any arbitrary $\n{\nu}$ can be expressed in terms of a black body distribution~\cite{Prigogine2015} 
with a frequency-dependent temperature $\Tbb{\nu}$ according to
\begin{equation}
\n{\nu} = \frac{\denp{\nu}}{e^{\frac{h\nu}{\kb\Tbb{\nu}}}-1} \,.
\end{equation}
The chemical potential~\eqref{eq:cpp} can thus be written as
\begin{equation}
\cpp{\nu} = h \nu
\bigg(1 - \frac{T}{\Tbb{\nu}}\bigg)\,,
\label{eq:cppBB}
\end{equation}
and, therefore, photons with frequency $\nu$ have a non-vanishing chemical potential 
if the temperature of the corresponding black body $\Tbb{\nu}$ is different from the temperature of the solution $T$
\rev{used as a reference}~\cite{wurfel1982,Ries1991,wurfel2005}.
This means, from a thermodynamic standpoint, 
that each set of photons with frequency $\nu$ plays the role of a thermostat with temperature $\Tbb{\nu}$.
\rev{The mismatch between the temperature $\Tbb{\nu}$ and the temperature $T$ of the main thermostat, i.e., the solvent,
defines the free energy carried by photons of frequency $\nu$
that can be used to maintain}
\transitions\ out of equilibrium~\cite{penocchio2021photo}.
If $\cpp{\nu} = 0$,  then $\Tbb{\nu} = T$ 
which thermodynamically means that the set of photons with frequency $\nu$ is in equilibrium with the solution
\rev{and does not carry any free energy that can be used to maintain
\transitions\ out of equilibrium.}


\subsection{Second Law\label{sub:SL0}}

The total free energy $\free$ of the solution is given by the sum of two contributions:
\begin{equation}
\free = \freec + \freep\,,
\label{eq:free_elem}
\end{equation}
where
\begin{equation}
\freec = 
{\sum_{\gs}}
{\sum_{\es}}
\big(\cp{\es} -\kb T \big)\conc{\Ze}
\label{eq:free_ch_elem}
\end{equation}
is the free energy contribution due to the molecular species~\cite{Fermi1956, Rao2016},
while 
\begin{equation}
\freep = \int \dd\nu \bigg[\underbrace{\cpp{\nu}\np - \kb T \denp{\nu}\ln\frac{ \denp{\nu} + \np}{\denp{\nu}}}_{\equiv\dfreep}\bigg]
\label{eq:free_ph_elem}
\end{equation}
is the free energy contribution due to the photons~\cite{Prigogine2015,penocchio2021photo},
with $\dfreep$ its free energy density per frequency.
By taking the time derivative of the total free energy~\eqref{eq:free_elem}
according to the rate equations~\eqref{eq:req} and~\eqref{eq:req_p}, 
we obtain the second law of thermodynamics in the form of
\begin{equation}
\dt \free = - T \epr + \pw = - T \eprq - T \eprp + \pw\,.
\label{eq:2law_elem}
\end{equation}
Here, $\epr$ is the total entropy production rate which is given by the sum of two contributions.
One is the entropy production rate $\eprq$ 
of the thermally induced \transitions~\eqref{eq:qrct} reading 
\begin{equation}
T \eprq = - 
\sum_{\gs,\gss}
\sum_{\ess>\es} 
(\cp{\ess} - \cp{\es}) 
\qcurr{\es}{\ess}
\geq 0\,,
\label{eq:eprq_elementary}
\end{equation}
which is non-negative because of the local detailed balance condition in Eq.~\eqref{eq:ldb_qrct}.
The other is the entropy production rate $\eprp$
of the photo-induced \transitions~\eqref{eq:prct} reading 
\begin{equation}
T \eprp =  
- 
\sum_{\gs,\gss}
\sum_{\ess>\es} 
(\cp{\ess} - \cp{\es} - \cpp{\fe}) 
\pcurr{\es}{\ess} 
\geq 0\,,
\label{eq:eprp_elementary}
\end{equation}
which is non-negative because of the local detailed balance condition in Eq.~\eqref{eq:ldb_prct}.
The last term featuring in Eq.~\eqref{eq:2law_elem}, i.e., $\pw$,
is the work rate quantifying the rate of free energy absorbed from the radiation, 
\begin{equation}
\pw =  \int\dd\nu\, \cpp{\nu} \excurr\,,
\label{eq:work_rate}
\end{equation}
maintaining \transitions\ out of equilibrium. 

\remark
If the chemical potential of the photons vanishes for every frequency $\nu$, 
namely, the temperature of the corresponding black body $\Tbb{\nu}$ is equal to the temperature of the solution $T$
for every frequency $\nu$,
also the work rate vanishes, 
namely, $\pw = 0$.
This physically means that the radiation is in equilibrium with the solution,
and all \transitions\ will eventually relax to equilibrium.
This can be formally seen from the second law~\eqref{eq:2law_elem} that boils down to $\dt \free = - T \eprq - T \eprp \leq 0$ if $\pw = 0$
together with the fact that $\free$ is lower bounded by its equilibrium value $\freeeq$~\cite{penocchio2021photo}.


\section{Coarse Graining of the Vibrational States:
Effective \Transitions\ Between Electronic States
\label{sec:cg_vibrational_states}}

We coarse grain here the dynamical description of Sub.~\ref{sub:el_kin} 
to derive a dynamical description in terms of 
species in specific electronic states, namely, $\{\Zg\}$,
regardless of their vibrational states.
Crucially, we construct the effective \transitions\ between electronic states
in such a way that they are thermodynamically consistent
and, furthermore, still satisfy a local detailed balance condition.


\subsection{Equilibration Within Electronic States\label{sub:cg1_ass}}

We assume that 
the photo-induced \transitions\ interconvert only 
vibrational states of different electronic states of the same species
(like in the example in Fig.~\ref{fig:basic}b).
\rev{This assumption is well justified for experiments where 
organic molecules are irradiated by UV-visible light.
Indeed, UV-visible light sources generate 
photons with frequencies~$\fe$ corresponding to the energy difference
between vibrational states of different electronic states of organic molecules,
i.e., $h \fe = |\stie{\egss{\gs}} - \stie{\egs{\gss}}|$.
Hence, we consider light sources that do not promote 
any photo-induced \transitions\ between 
vibrational states of the same electronic state}~\footnote{\rev{
Note that photons with frequencies~$\fevib$ corresponding to the energy difference 
between vibrational states of the same electronic state, 
i.e., $h \fevib = |\stie{\egss{\gs}} - \stie{\egs{\gs}}|$, 
can still be present even in experiments where organic molecules 
are irradiated by UV-visible light.
Indeed, they may be generated by the solution itself
as any body at certain temperature can emit light. 
However, their density is negligible compared to the density of photons with frequency in the UV-visible spectrum.
Furthermore, these photons are in equilibrium with the solution itself, i.e., $\cpp{\fevib} = 0$,
and the corresponding photo-induced \transitions\ thus behave in practice like thermally induced ones 
(compare the local detailed balance for photo-induced \transitions\ in Eq.~\fref{eq:ldb_prct} when $\cpp{\nu} = 0$
with the local detailed balance for thermally induced \transitions\ in Eq.~\fref{eq:ldb_qrct})}}:
\begin{equation}
\pcurr{\es}{\egss{\gs}} = 0
\label{eq:no_vibrational_photo_induced_transition}
\end{equation}
for all $\es$ and $\egss{\gs}$.
We further assume
that
the \transitions\ between vibrational states of the same electronic state
occur on a much shorter timescale than 
the \transitions\ 
between vibrational states of different electronic states.
This assumption, often called \textit{vibrational relaxation}, 
\rev{is well justified for experiments where organic molecules 
are irradiated by UV-visible light:
vibrational transitions within the same electronic states occur on a time scale of the order of $10^{-12}$~seconds,
while transitions between different electronic states occur on a time scale of the order of $10^{-8}$~seconds
(in the absence of phosphorescent states)~\cite{Balzani2014}.}
Hence, for any $\gs$, on the time scale at which the concentration $\conc{\Zg}$ remains almost constant, 
the concentrations $\{\conc{\Ze}\}$ dramatically change
and 
the thermally induced \transitions~\eqref{eq:qrct} between any $\es$ and $\egss{\gs}$ equilibrate.
This has two main implications.

From a dynamical point of view, 
\begin{equation}
{\qflux{\es}{\egss{\gs}}} = {\qflux{\egss{\gs}}{\es}}
\label{eq:dynamical_eq}
\end{equation}
at all times for every pair of vibration states $\es$ and $\egss{\gs}$
or, equivalently, ${\qcurr{\es}{\egss{\gs}}} = 0$.
Hence, there exist 
\begin{equation}
\prob{\es}{\gs} = 
\frac{e^{-\frac{\stcp{\es}}{\kb T}}}
{\partition{\gs}}\,,
\label{eq:prob_i_in_alpha}
\end{equation}
with $\partition{\gs} = \sum_{\es} \exp({-{\stcp{\es}}/{\kb T}})$,
named hereafter the population of the vibrational state $\es$,
such that
\begin{equation}
\conc{\Ze} = \prob{\es}{\gs} \conc{\Zg}\,.
\label{eq:conc_v_eq}
\end{equation}
This is a direct consequence of the fact that the \transitions\ between vibration states of the same electronic state
define a detailed balanced linear network.  
Indeed, by using Eq.~\eqref{eq:conc_v_eq} in the expression for the chemical potentials~\eqref{eq:cp} 
together with the local detailed balance condition~\eqref{eq:ldb_qrct}, 
one can easily verify that Eq.~\eqref{eq:dynamical_eq} is satisfied. 

From a thermodynamic point of view, 
\begin{equation}
\cp{\es} = \cp{\egss{\gs}} 
\label{eq:equilibration_thermo}
\end{equation}
at all times for every pair of vibration states $\es$ and $\egss{\gs}$.
This allows us to define the chemical potential of~$\gs$ as
\begin{equation}
\cp{\gs} \equiv \cp{\es} = \underbrace{\stcp{\es} +\kb T\ln \prob{\es}{\gs}}_{\equiv\stcp{\gs} } + \kb T\ln \conc{\Zg} \,,
\label{eq:cp_cg1}
\end{equation}
by using Eq.~\eqref{eq:conc_v_eq} in Eq.~\eqref{eq:cp} and Eq.~\eqref{eq:equilibration_thermo}.
Note that the standard chemical potential $\stcp{\gs}$ is independent of $\es$ 
because of Eq.~\eqref{eq:prob_i_in_alpha}
implying that $\stcp{\gs} = -\kb T \ln \partition{\gs}$.

\remark 
The interconversions between vibrational states of the same electronic state would not equilibrate
if they could occur via photon-induced \transitions\ too.
Indeed, the interplay between thermally induced \transitions\ (promoted by the thermostat at temperature $T$)
and photo-induced \transitions\ (promoted by photons playing the role of other thermostats at temperatures $\{\Tbb{\nu}\}$)
would lead to a nonequilibrium steady state~\cite{penocchio2021photo}.
If this were the case, a thermodynamically consistent coarse-grained description in terms of 
\transitions\ between electronic states only could not be derived~\cite{Esposito2012}.


\subsection{Kinetics\label{sec:cg1:kinetics}}

\begin{figure}
    \centering
    \includegraphics[width=0.49\textwidth]{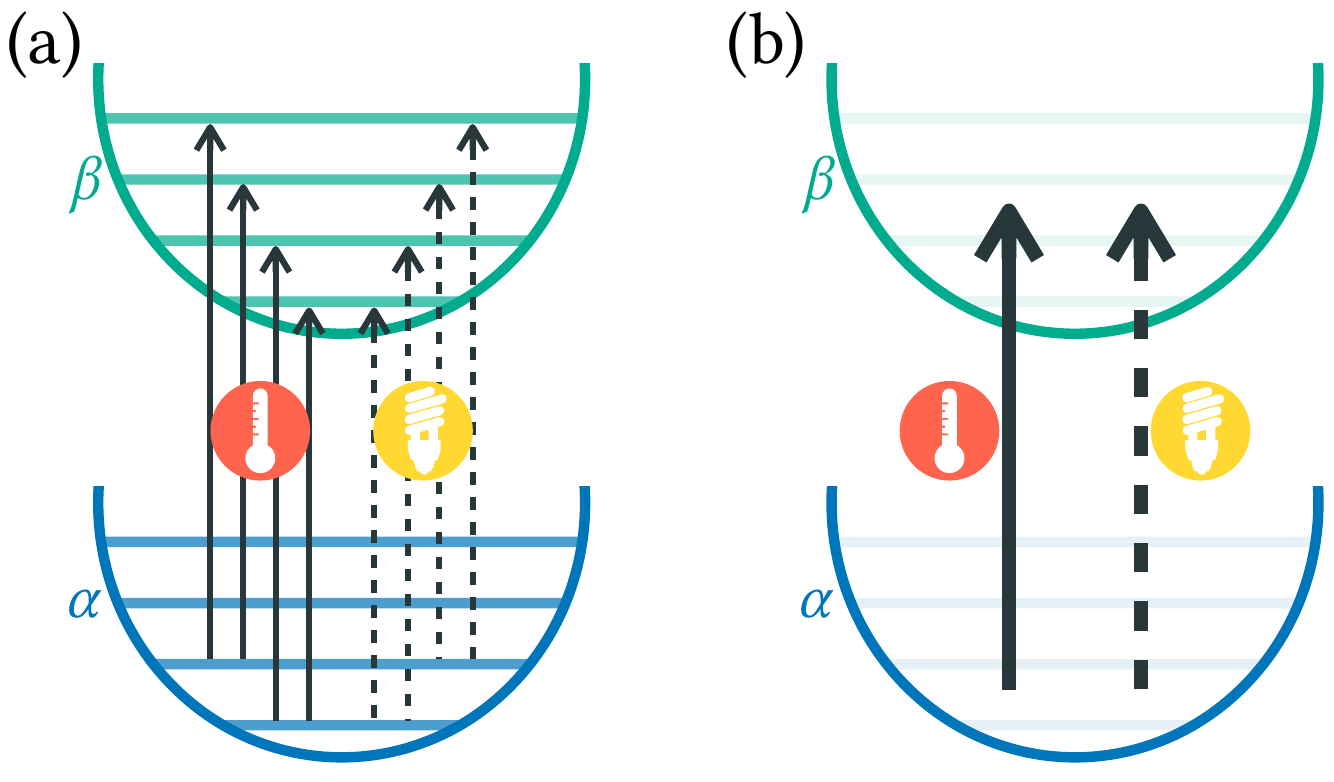}
    \caption{
    Thermally induced (arrows close to a thermometer) and 
    photo-induced (dashed arrows close to a bulb) 
    \transitions\
    between two electronic states of the same species~$\gs$ and~$\gss$.
    (a) Thermally and photo-induced \transitions\ 
    between (some of) the vibrational states (represented by horizontal lines) 
    of the electronic states~$\gs$ and~$\gss$. 
    (b) All thermally induced (resp. photo-induced) \transitions\ are combined together to form 
    an effective thermally induced (resp. photo-induced) \transition\ 
    between the electronic states of the same species~$\gs$ and~$\gss$.
    Each \transition\ has always a backward counterpart, even if it is not represented.
    \label{fig:cg1}}
\end{figure}

An effective thermally induced \transition\ from the electronic state~$\gs$ to the electronic state~$\gss$ 
results from the combination of
all thermally induced \transitions\ from the vibrational states $\{\es\}$ to the vibrational states $\{\ess\}$
(see Fig.~\ref{fig:cg1}).
Thus, we can define the flux $\qflux{\gs}{\gss}$ as 
\begin{equation}
\qflux{\gs}{\gss} \equiv
\sum_{\es,\,\ess} 
\qflux{\es}{\ess}
\label{eq:flux_q_cg1}
\end{equation}
which can be written as
\begin{equation}
\qflux{\gs}{\gss} 
= 
\kq{\gs}{\gss}\conc{\Zg}\,,
\label{eq:flux_q_e}
\end{equation}
by using Eqs.~\eqref{eq:flux_q} and~\eqref{eq:conc_v_eq} and defining the kinetic constant 
\begin{equation}
\kq{\gs}{\gss} \equiv 
\sum_{\es,\,\ess}  
\kq{\es}{\ess} \, \prob{\es}{\gs}\,.
\end{equation}
The thermally induced \transition\ from $\gs$ to $\gss$ 
and its backward counterpart from $\gss$ to $\gs$ can be represented by the following chemical equations 
\begin{subequations}
\begin{align}
{\Zg} &\ch{->} {\Zgg}\,,\\
{\Zgg} &\ch{->} {\Zg} \,,
\end{align}
\label{eq:qrct_cg1}
\end{subequations} 
and the corresponding net current reads
\begin{equation}
\qcurr{\gs}{\gss} = \qflux{\gs}{\gss}  - \qflux{\gss}{\gs}\,.
\label{eq:cg1currq}
\end{equation}

\remark
These effective thermally induced \transitions\ 
can interconvernt the electronic states of the same species 
(like in Fig.~\ref{fig:cg1})
as well as the electronic states of different species.

Similarly, 
an effective photo-induced \transition\ from the electronic state $\gs$ to the electronic state $\gss$
of the same species
results from the combination of
all photo-induced \transitions\ from the vibrational states $\es$ to the vibrational states $\ess$
(see Fig.~\ref{fig:cg1}).
Here, we assume that when the electronic state energies, 
defined as $\stie{\gs} = \sum_{\es}\stie{\es}\prob{\es}{\gs}$ and $\stie{\gss} = \sum_{\ess}\stie{\ess}\prob{\ess}{\gss}$, 
satisfy $\stie{\gss} > \stie{\gs}$, 
then the only possible photo-induced \transitions\ from $\gs$ to $\gss$ (resp. $\gss$ to $\gs$) 
are absorption (resp. emission) \transitions . 
This physically means that there are no populated vibrational states of $\gs$ with 
energies higher
than those of the populated vibrational states of $\gss$,
which is a valid assumption at room temperature.
We can hence define the density fluxes of absorption and emission
between the electronic states $\gs$ and $\gss$ (with $\stie{\gss} > \stie{\gs}$)
promoted by photons with frequency $\nu$ as
\begin{subequations}
\begin{align}
\pdfluxa{\gs}{\gss}{\nu} &\equiv 
\sum_{\es,\,\ess}  
\ddirac{\nu}{\fe} \pfluxa{\es}{\ess} \,,\\
\pdfluxe{\gss}{\gs}{\nu} & \equiv 
\sum_{\es,\,\ess}  
\ddirac{\nu}{\fe} \pfluxe{\ess}{\es}\,,
\end{align}
\label{eq:flux_p_cg1}
\end{subequations}
respectively. 
They can be written as
\begin{subequations}
\begin{align}
\pdfluxa{\gs}{\gss}{\nu} & = 
\dkpa{\gs}{\gss}{\nu}\,\np\conc{\Zg}
\,,\\
\pdfluxe{\gss}{\gs}{\nu} & = 
\big(\dkpe{\gss}{\gs}{\nu} + \dkpse{\gss}{\gs}{\nu}  \,\np\big)\conc{\Zgg}
\,,
\end{align}
\end{subequations}
by using Eqs.~\eqref{eq:flux_p} and~\eqref{eq:conc_v_eq} and 
defining the frequency-dependent kinetic constants
\begin{subequations}
\begin{align}
\dkpa{\gs}{\gss}{\nu} & = 
\sum_{\es\,,\ess} 
\ddirac{\nu}{\fe} \kpa{\es}{\ess}\, \prob{\es}{\gs}
\,,\\
\dkpe{\gss}{\gs}{\nu} & = 
\sum_{\es,\,\ess} 
\ddirac{\nu}{\fe} \kpe{\ess}{\es}\, \prob{\ess}{\gss}
\,,\\
\dkpse{\gss}{\gs}{\nu} & = 
\sum_{\es,\,\ess}
\ddirac{\nu}{\fe} \kpse{\ess}{\es}\, \prob{\ess}{\gss}
\,.
\end{align}
\end{subequations}
These \transitions\ can be represented by the following chemical equations
\begin{subequations}
\begin{align}
\Zg + \p &\ch{->} \Zgg\,, \\
\Zgg &\ch{->} \Zg + \p\,, \\
\Zgg + \p &\ch{->} \Zg + 2\p\,,
\end{align}
\label{eq:prct_cg1}
\end{subequations}
and the corresponding net density current reads
\begin{equation}
\pdcurr{\gs}{\gss}{\nu} = \pdfluxa{\gs}{\gss}{\nu} - \pdfluxe{\gss}{\gs}{\nu} \,.
\label{eq:cg1currp}
\end{equation}
Note that $\pdcurr{\gs}{\gss}{\nu} = \pdfluxe{\gs}{\gss}{\nu} - \pdfluxa{\gss}{\gs}{\nu}$ 
if $\stie{\gss} < \stie{\gs}$.

\remark
These effective photo-induced \transitions\ 
can interconvernt the electronic states of the same species only
(like in Fig.~\ref{fig:cg1}).


\subsection{Rate Equations\label{sub:cg1_kin}}

The concentration of species in a specific electronic state, i.e., $\conc{\Zgg}$, 
is affected by both the effective thermally  and photo-induced \transitions\ 
and follows
\begin{equation}
\dt\conc{\Zgg} = 
\sum_{\substack{\gs \\ (\gs \neq \gss)}}
\bigg\{
\qcurr{\gs}{\gss} 
+ 
\underbrace{\int\dd\nu\,\pdcurr{\gs}{\gss}{\nu}}_{= \pcurr{\gs}{\gss} }
\bigg\}\,,
\label{eq:req_cg1}
\end{equation}
as it can be verified by
i) using Eq.~\eqref{eq:req} in $\dt\conc{\Zgg} = \sum_{\ess\in\gss}\dt\conc{\Zee}$,
ii) the definitions in Sec.~\ref{sec:cg1:kinetics},
and iii) recalling that $\qcurr{\es}{\egss{\gs}} = 0$ and $\pcurr{\es}{\egss{\gs}} = 0$ for all $\es$ and $\egss{\gs}$.

On the other hand, 
the rate equation~\eqref{eq:req_p} for the density of photons with frequency~$\nu$ can be simply rewritten as
\begin{equation}
\dt\n{\nu} = -
\sum_{\gss>\gs}
\pdcurr{\gs}{\gss}{\nu} +  \excurr \,,
\label{eq:req_p_cg1}
\end{equation}
where
we ordered the electronic states, without loss of generality, in such a way that $\gss > \gs$ if $\stie{\gss} > \stie{\gs}$.
Equation~\eqref{eq:req_p_cg1} can be simply obtained by
using the definitions in Sec.~\ref{sec:cg1:kinetics} 
and recalling that $\pcurr{\es}{\egss{\gs}} = 0$ for all $\es$ and $\egss{\gs}$.


\subsection{Local Detailed Balance\label{sub:LDB1}}

We show now that the fluxes between electronic states 
defined in Eqs.~\eqref{eq:flux_q_cg1} and~\eqref{eq:flux_p_cg1} still satisfy a local detailed balance condition.

For thermally induced \transitions , we can write
\begin{equation}
\frac{\qflux{\gs}{\gss}}{\qflux{\gss}{\gs}} = 
\frac{
\sum_{\es,\,\ess}
\qflux{\ess}{\es} e^{-\frac{\cp{\ess} - \cp{\es}}{\kb T}}
}{
\sum_{\es,\,\ess} 
\qflux{\ess}{\es}
}\,,
\end{equation}
by using the definition in Eq.~\eqref{eq:flux_q_cg1} 
and Eq.~\eqref{eq:ldb_qrct}.
However, the equilibration of the thermally induced \transitions\ within an electronic state implies that 
the chemical potentials $\cp{\ess}$ and $\cp{\es}$ are equal to 
the chemical potentials of the corresponding electronic state
$\cp{\gss}$ and $\cp{\gs}$
(see Eq.~\eqref{eq:cp_cg1}).
Hence, we obtain the following local detailed balance condition:
\begin{equation}
\kb T \ln \frac{\qflux{\gs}{\gss}}{\qflux{\gss}{\gs}} = - \big(\cp{\gss} - \cp{\gs} \big)\,.
\label{eq:ldb_qrct_cg1}
\end{equation}

For photo-induced \transitions , we can write
\begin{equation}\small
\frac{\pdfluxa{\gs}{\gss}{\nu}}{\pdfluxe{\gss}{\gs}{\nu}} = 
\frac{
\sum_{\es,\,\ess}
\ddirac{\nu}{\fe} \pfluxe{\ess}{\es} e^{-\frac{\cp{\ess} - \cp{\es} - \cpp{\nu}}{\kb T}}
}{
\sum_{\es,\,\ess}
\ddirac{\nu}{\fe} \pfluxe{\ess}{\es}
}\,,
\end{equation}
if $\stie{\gss} > \stie{\gs}$,
by using the definition in Eqs.~\eqref{eq:flux_p_cg1} 
and Eq.~\eqref{eq:ldb_prct}.
As for the thermally induced \transitions , 
$\cp{\ess} = \cp{\gss}$ $\forall \ess$ 
and $\cp{\es}= \cp{\gs}$ $\forall \es$, 
and, therefore, we obtain the following local detailed balance condition:
\begin{equation}
\kb T \ln\frac{\pdfluxa{\gs}{\gss}{\nu}}{\pdfluxe{\gss}{\gs}{\nu}} = - \big(\cp{\gss} - \cp{\gs} -  \cpp{\nu} \big)\,.
\label{eq:ldb_prct_cg1}
\end{equation}

\remark
As explained in Subs.~\ref{sub:ldb},
photons with frequency~$\nu$ correspond to a black body with temperature $\Tbb{\nu}$,
namely, a thermostat with temperature $\Tbb{\nu}$.
This allows us to interprete the $\nu$-dependence in Eq.~\eqref{eq:ldb_prct_cg1},
in terms of the local detailed balance condition resolving 
the specific thermostat promoting the \transition\ from $\gs$ to $\gss$.
If the light source were an actual black body at temperature $\TTbb$,
the chemical potential of the photons~\eqref{eq:cppBB} would be frequency independent 
since $\Tbb{\nu} = \TTbb$ for all $\nu$. 
A local detailed balance condition would then hold also for the fluxes integrated over the frequency, namely,
\begin{equation}
\kb T \ln\frac{
\int \dd\nu\,
\pdfluxa{\gs}{\gss}{\nu}
}{
\int \dd\nu\,
\pdfluxe{\gss}{\gs}{\nu}} = - \big(\cp{\gss} - \cp{\gs} -  \cppbb \big)\,
\end{equation}
with $\cppbb = h \nu (1 - {T}/{\TTbb})$.


\subsection{Second Law\label{sub:SL1}}

The second law of thermodyanics~\eqref{eq:2law_elem}
still holds after the coarse graining of the vibrational states.
Indeed, all thermodynamic quantities introduced in Subs.~\ref{sub:SL0} 
can be expressed 
in terms of species in specific electronic states regardless of their vibrational states and of the density of photons.
This is a direct consequence of the equilibration of the vibrational states of the same electronic state~\cite{Esposito2012} as we show in the following.

First,
the free energy contribution due to the molecular species~\eqref{eq:free_ch_elem} boils down 
\begin{equation}
\freec = \sum_{\gs}\big(\cp{\gs} -\kb T \big)\conc{\Zg}
\label{eq:free_ch_cg1}
\end{equation}
because of Eq.~\eqref{eq:cp_cg1}.
Second,
the entropy production rate of the 
thermally induced~\eqref{eq:eprq_elementary} 
and photo-induced~\eqref{eq:eprp_elementary} transitions become
\begin{subequations}\small
\begin{align}
T \eprq &= - 
\sum_{\gss>\gs}(\cp{\gss} - \cp{\gs}) \qcurr{\gs}{\gss} \geq 0\,,
\label{eq:eprq} \\
T \eprp &=  \int\dd\nu\, \Big[\underbrace{
- 
\sum_{\gss>\gs}(({ \cp{\gss} - \cp{\gs} - \cpp{\nu}}) \pdcurr{\gs}{\gss}{\nu}
}_{\equiv T\deprp\geq 0} \Big]\geq 0\,,
\label{eq:eprp}
\end{align}\label{eq:epr_cg1}%
\end{subequations}
by using
Eqs.~\eqref{eq:no_vibrational_photo_induced_transition},~\eqref{eq:dynamical_eq},~\eqref{eq:cp_cg1},~\eqref{eq:flux_q_cg1} and~\eqref{eq:flux_p_cg1}.
Note that $ \eprq $ and $\deprp$ in Eq.~\eqref{eq:epr_cg1} are 
non-negative because of the local detailed balance conditions in Eqs.~\eqref{eq:ldb_qrct_cg1} and~\eqref{eq:ldb_prct_cg1}.
Third,
the free energy contribution due to the photons and the work rate 
are still given 
in Eqs.~\eqref{eq:free_ph_elem} and~\eqref{eq:work_rate}, 
respectively.
Finally, one can easily verify (by direct substitution) that 
the thermodynamic quantities in 
Eqs.~\eqref{eq:free_ph_elem},~\eqref{eq:work_rate},~\eqref{eq:free_ch_cg1}, and~\eqref{eq:epr_cg1}
satisfy the second law of thermodyanics in Eq.~\eqref{eq:2law_elem}.


\section{Coarse Graining of the Excited States:
\Isomerization\ \Reactions\ 
\label{sec:HDM}}

\begin{figure}
    \centering
    \includegraphics[width=0.49\textwidth]{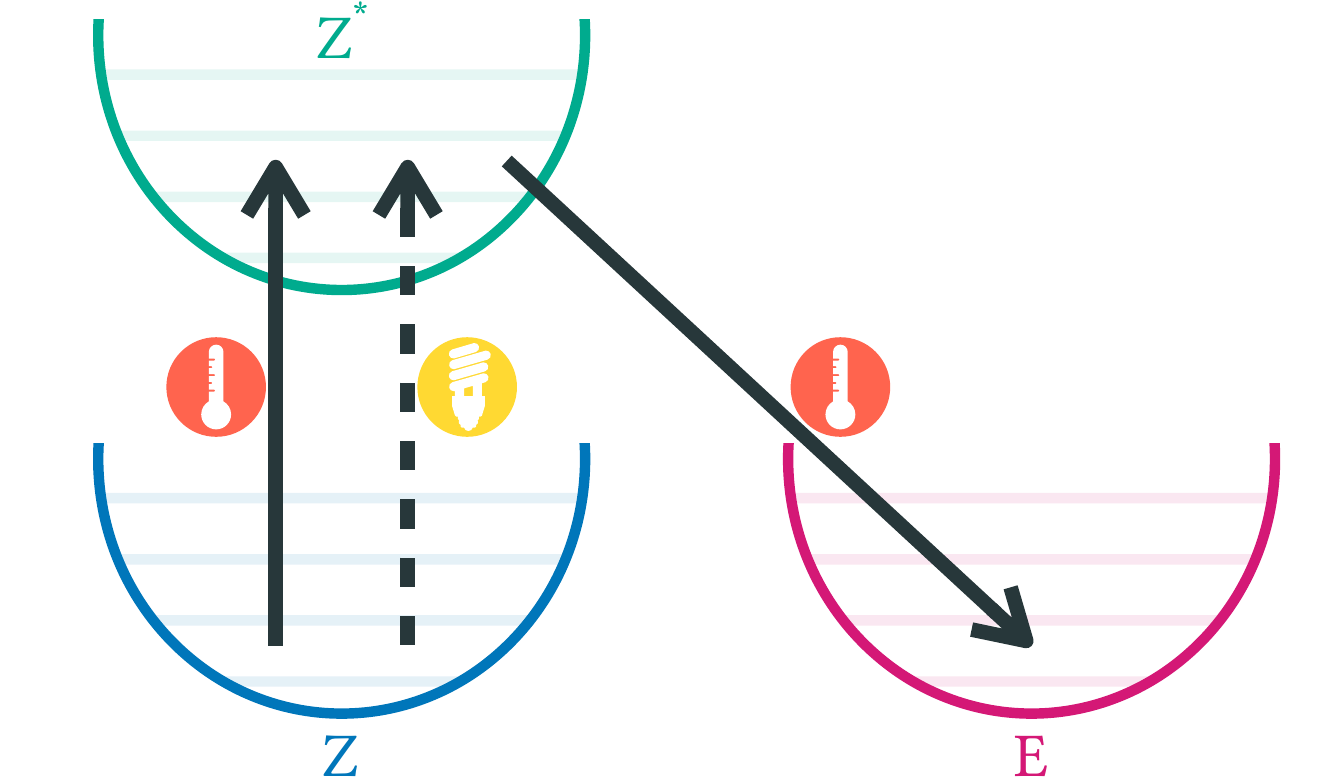}
    \caption{
    A species in the electronic ground state~$\cZ$ can be interconverted into its electronic excited state~$\cZZ$
    via a thermally induced (vertical arrow close to a thermometer) 
    and a photo-induced (vertical dashed arrow close to a bulb)
    \transition . 
    The electronic excited state~$\cZZ$ can also be interconverted 
    via a thermally induced (oblique arrow close to a thermometer) \transition\ 
    into the the electronic ground state of a different species~$\cE$.
    Chemically speaking, the species~$\cZ$ and~$\cE$ are different isomers.
    \rev{Recall that no photo-induced transition can interconvert the ground state of species~$\cE$ 
    into the excited state of a different species~$\cZZ$ as a result of the Franck-Condon principle~\cite{Balzani2014}.
    Note that each \transition\ has always a backward counterpart, even if it is not represented.}
    \label{fig:HDM}}
\end{figure}

We coarse grain here the dynamical description of Sub.~\ref{sub:cg1_kin} 
to derive a dynamical description in terms 
of ground-state species only.
Some of the resulting effective \transitions\ 
interconvert a ground-state species into a different one,
thus constituting an \isomerization .
We will refer to them as \isomerization\ \reactions . 
Other effective \transitions\ 
interconvert the ground state of a species back into itself 
without realizing any transformation.
We will refer to them as \futile\ \reactions .
Crucially, we construct all these \reactions\
in such a way that they are thermodynamically consistent
and, furthermore, still satisfy a local detailed balance condition.

We focus here on the case illustrated in Fig.~\ref{fig:HDM},
where we replaced the general labels of the electronic states used until now 
(i.e., $\gs$ and $\gss$) with specific labels.
The \reactions\ resulting from the coarse graining are shown in Fig.~\ref{fig:HDM_effrct}.
Two of them are \isomerization\ \reactions\ (Fig.~\ref{fig:HDM_effrct}a-b) 
while the other two are \futile\ \reactions\ (Fig.~\ref{fig:HDM_effrct}c-d).

\remark
From now on, 
we will refer to any \transition\ involving ground-state species 
(independently of whether it is an effective transition or not)
as a \reaction .


\subsection{Relaxation of the Electronic Excited State}

We assume that the evolution of the excited-state species $\cZZ$ is much faster than 
the evolution of the ground-state species $\cZ$ and $\cE$\rev{, 
reflecting the fact that $\cZZ$ is generally short-lived.
This is a valid assumption for experiments where organic molecules 
are irradiated by UV-visible light since the lifetime of excited-state species is of the order of $10^{-8}$~seconds
in the absence of long-lived phosphorescent states that we are not considering here~\cite{Balzani2014}.}
Hence, on the time scale at which 
the concentrations $\conc{\cZ}$ and $\conc{\cE}$ remain almost constant, 
the concentration $\conc{\cZZ}$ dramatically changes
and 
relaxes to the corresponding steady state whose concentration reads
\begin{equation}\small
\ssconc{\cZZ} = 
\frac{
\kq{\cZ}{\cZZ} \conc{\cZ}
+ \int \dd\nu\, \dkpa{\cZ}{\cZZ}{\nu} \np \conc{\cZ}
+ \kq{\cE}{\cZZ} \conc{\cE}
}{
\DD{\cZZ}
}\,,
\label{eq:ss_hdm}
\end{equation}
with 
\begin{equation}\small
\DD{\cZZ} = 
\kq{\cZZ}{\cZ}
+ \int \dd\nu\Big(\dkpe{\cZZ}{\cZ}{\nu} + \dkpse{\cZZ}{\cZ}{\nu}  \,\np\Big)
+ \kq{\cZZ}{\cE}\,.
\label{eq:DD}
\end{equation}
Equation~\eqref{eq:ss_hdm} is obtained by imposing $\dt \conc{\cZZ} = 0$ in Eq.~\eqref{eq:req_cg1}.

\remark
While in Subs.~\ref{sub:cg1_ass} we assumed the equilibration of the vibrational states,
here we are solely using a quasi-steady-state approximation.
Nevertheless, we can construct \reactions\ between ground-state species
that still satisfy a local detailed balance condition as we prove in the following.


\subsection{Cycle Kinetics\label{sec:cghdm:kinetics}}

\begin{figure}[t]
    \centering
    \includegraphics[width=0.49\textwidth]{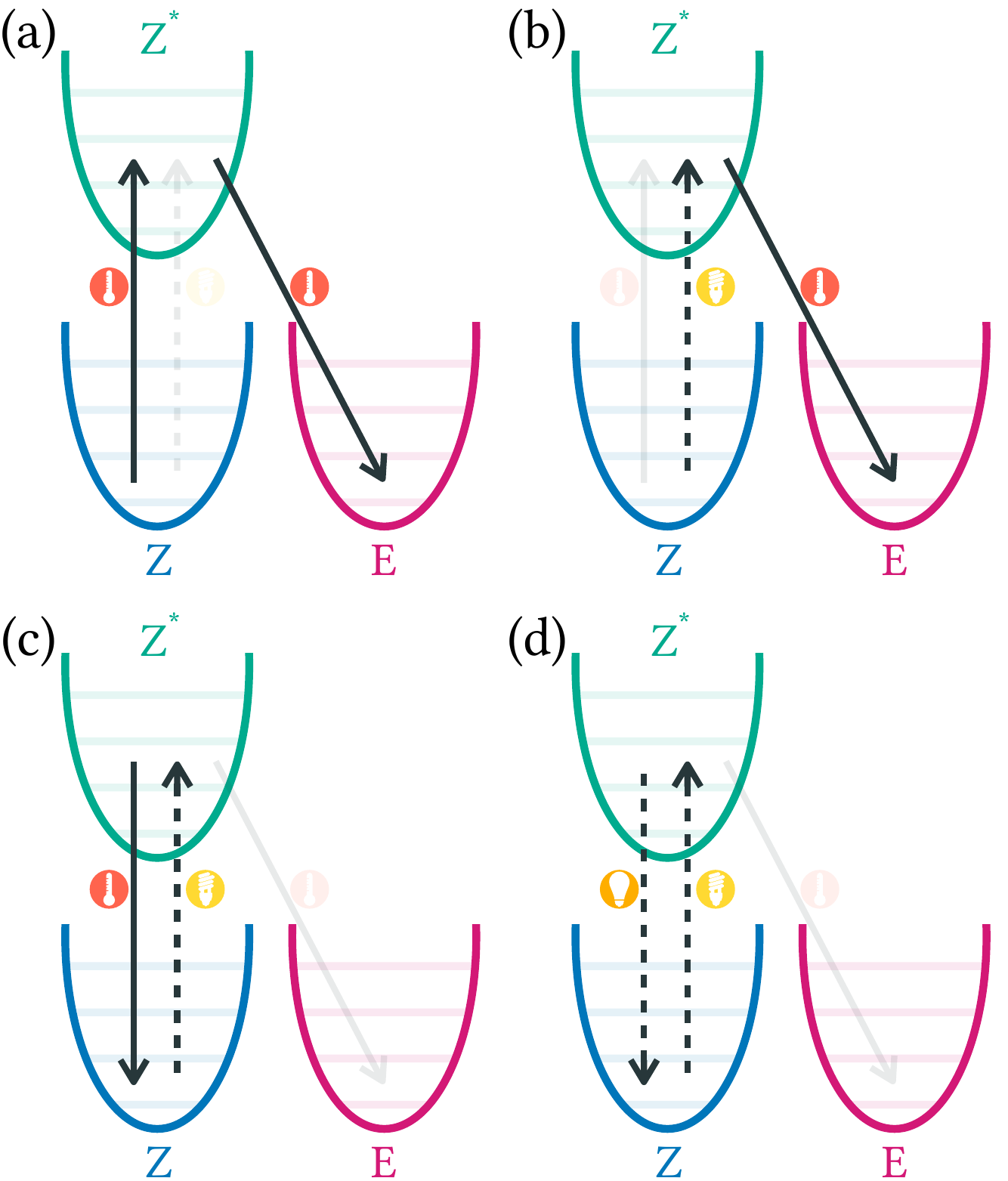}
    \caption{
    Cycles (represented by sequences of arrows) for the case in Fig.~\ref{fig:HDM}.
    Each cycle can be run in two directions: 
    one direction is illustrated by the arrows while the other is the opposite direction.
    \rev{By following the sequences of \transitions\ of each cycle,
    the excited-state species $\cZZ$ is never interconverted
    as it is always produced by one \transition\ ad consumed by another.}
    Note that cycle (a) can been obtained by running
    cycle (c) in the opposite direction with respect to the one determined by the arrows
    followed by cycle (b) 
    (namely, cycle (a) is linearly dependent on cycle (b) and (c)).
    Note also that cycle (d) results from a sequence of two photo-induced \transitions\ involving photons with different frequency 
    (represented by \rev{different bulbs of different colors}).
    \label{fig:HDM_cycles}}
\end{figure}

\begin{figure}
    \centering
    \includegraphics[width=0.49\textwidth]{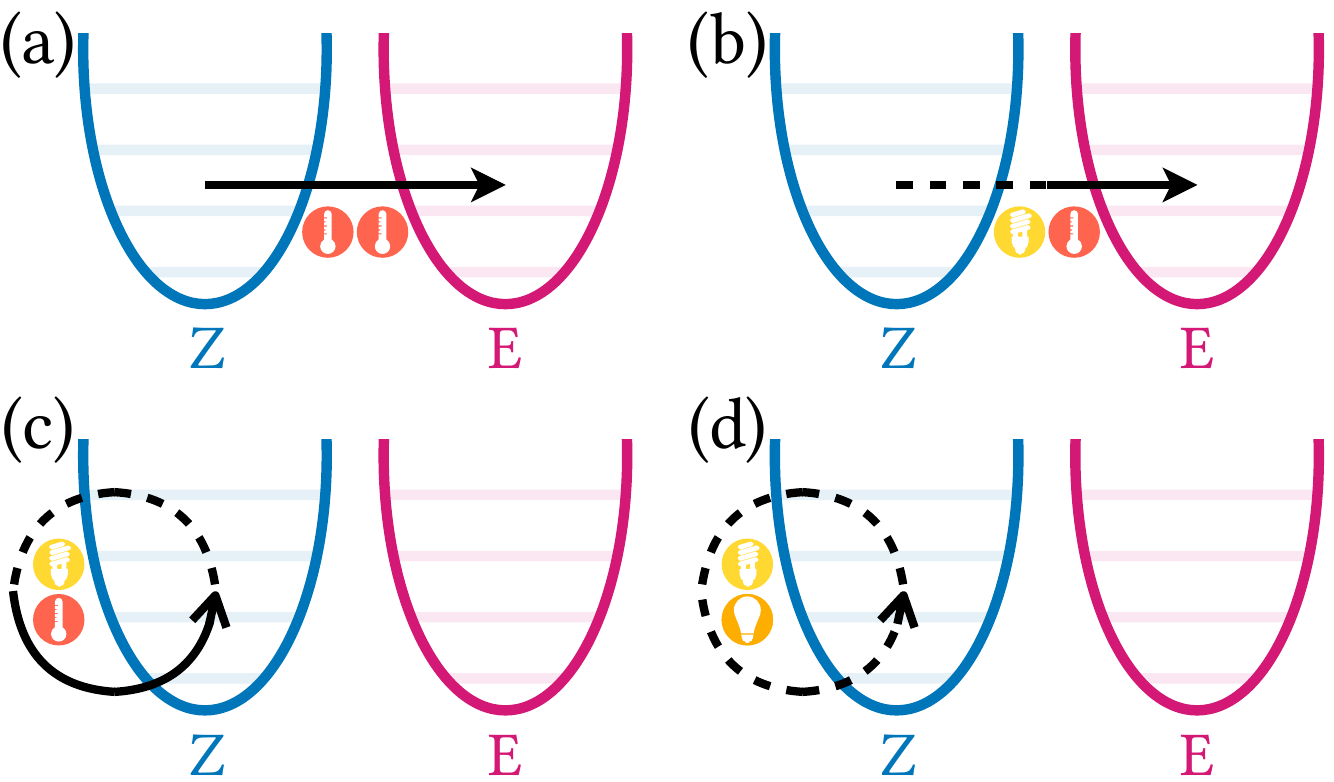}
    \caption{
    \Reactions\ between ground-state species corresponding to the cycles in Fig.~\ref{fig:HDM_cycles}.
    Each \reaction\ has always its backward counterpart even if it is not represented.
    Thermometers and bulbs specify whether the underlying \transitions\ are 
    thermally  or photo-induced or a combination of both.
    (a)~\Isomerization\ \reaction\ of $\cZ$ into $\cE$ 
    via a sequence of only thermally induced \transitions .
    (b)~\Isomerization\ \reaction\ of $\cZ$ into $\cE$ 
    via a sequence of both photo- and thermally induced \transitions .
    (c)~\Futile\ \reaction\ of $\cZ$ into itself 
    via a sequence of both photo- and thermally induced \transitions .
    Its net effect is the interconversion of photons into heat.
    (d)~\Futile\ \reaction\ of $\cZ$ into itself 
    via a sequence of only photo-induced \transitions .
    Its net effect is the interconversion of photons with one frequency into photons with a different frequency.
    \rev{Recall that each \reaction\ is defined by the sequence of \transitions\ in Fig.~\ref{fig:HDM_cycles} 
    that involves the excited-state species $\cZZ$.}
    \label{fig:HDM_effrct}}
\end{figure}

\rev{This time scale separation between excited-state species and ground-state species further implies that, 
as the concentrations of the ground-state species $\conc{\cZ}$ and $\conc{\cE}$ slowly evolve,
the concentration of the excited-state species $\conc{\cZZ}$ follows 
the steady-state concentration $\ssconc{\cZZ}$ in Eq.~\eqref{eq:ss_hdm}.
The dynamics can therefore be described in terms of so-called cycles~\cite{Avanzini2020b}.
In this context, a cycle is defined as a
sequence of \transitions\ that, 
upon completion, 
does not interconvert the excited-state species ${\cZZ}$
(which is and remains at steady state),
but only interconverts the ground-state species 
${\cZ}$ and ${\cE}$
(which are slowly evolving)~\footnote{\rev{
Intuitively, the term cycle might be associated with a sequence of \transitions\ that, upon completion, 
does not interconvert any species, unlike in our definition. 
However, due to the time scale separation, 
the concentrations of ground-state species evolve so slowly compared to the excited-state species
that they can be treated as constant.
As a result, a sequence of \transitions\ that, upon completion, does not interconvert the excited-state species 
can still be regarded as a cycle, since the ground-state species are effectively not interconverted either
given their much slower evolution.
For a more formal discussion of this point, see Ref.~\cite{Avanzini2020b}}}.
Its net effect is an (\isomerization\ or \futile ) \reaction\ 
between ground-state species determining their slow evolution
(even thought the underlying sequence of \transitions\ involves the excited-state species).
For the case in Fig.~\ref{fig:HDM},
the cycles and the corresponding (\isomerization\ or \futile ) \reactions\
are illustrated in Figs.~\ref{fig:HDM_cycles} and~\ref{fig:HDM_effrct}, respectively.}


We stress that (sets of linearly independent) cycles have been already used 
to construct thermodynamically consistent coarse-grained descriptions 
of chemically driven molecular systems~\cite{Wachtel2018, Avanzini2020b, Avanzini2023} 
as well as photo-induced \transitions~\cite{penocchio2021photo}.
However, the resulting \reactions\ do not, in general, satisfy
a local detailed balance condition~\cite{Wachtel2018, penocchio2021photo}.
Indeed, a local detailed balance condition can only be restored 
by using a conformal set of cycles~\cite{bilancioni2024elementaryfluxmodescrn}
which unfortunately might change during the dynamics 
since it is defined by the values of all net currents.

Here, 
we construct \reactions\ between ground-state species which still satisfy a local detailed balanced condition
by considering all possible cycles (independently of whether they are linearly independent or not).
This can be done because the thermally  and photo-induced \transitions\ we consider are 
linear in the concentrations $\conc{\cZ}$, $\conc{\cZZ}$, and $\conc{\cE}$
and, therefore, we can exploit a modified version of the matrix-tree theorem~\cite{Hill1966, HILL197733} 
already used in Ref.~\cite{LiangBusiello2024}.

We start by recognizing that the cycles in Fig.~\ref{fig:HDM_cycles} constitute all possible cycles 
of the case in Fig.~\ref{fig:HDM}.
Each cycle can be run in two directions: 
one is the direction illustrated by the arrows in Fig.~\ref{fig:HDM_cycles} 
while the other is the opposite direction.
One direction defines the forward \reaction\ between ground-state species, 
while the other defines the backward one.
According to the modified matrix-tree theorem~\cite{Hill1966, HILL197733} used in Ref.~\cite{LiangBusiello2024},
the corresponding (density) fluxes
are given by the product of the (density) fluxes of the underlying \transitions\ along the cycle 
divided by $\DD{\cZZ}$ in Eq.~\eqref{eq:DD} and the excited-state concentration~$\conc{\cZZ}$
(which will cancel out with an equivalent term featuring the numerator).

Hence, 
the fluxes of the cycle in Fig.~\ref{fig:HDM_cycles}a 
(or, equivalently, of the \isomerization\ \reaction\ in Fig~\ref{fig:HDM_effrct}a) 
read
\begin{subequations}\small
\begin{align}
\qcflux{\cZ}{\cE}{\cZZ} & = 
\frac{
\kq{\cZ}{\cZZ}\conc{\cZ}  \kq{\cZZ}{\cE}
}{
\DD{\cZZ}
}\,,\\
\qcflux{\cE}{\cZ}{\cZZ} & =
\frac{
\kq{\cE}{\cZZ}\conc{\cE}  \kq{\cZZ}{\cZ}
}{
\DD{\cZZ}
} \,,
\end{align}
\label{eq:flux_qZtoEvZZ}
\end{subequations}
while the corresponding net current is given by 
\begin{equation}\small
\qccurr{\cZ}{\cE}{\cZZ}  = 
\qcflux{\cZ}{\cE}{\cZZ}
-
\qcflux{\cE}{\cZ}{\cZZ}
\,.
\label{eq:currQ:Z_ZZ_E}
\end{equation}

The density fluxes of absorption and emission of the cycle in Fig.~\ref{fig:HDM_cycles}b 
(or, equivalently, of the \isomerization\ \reaction\ in Fig~\ref{fig:HDM_effrct}b)
 read
\begin{subequations}\small
\begin{align}
\pcdfluxa{\cZ}{\cE}{\cZZ}{\nu} & = 
\frac{
\dkpa{\cZ}{\cZZ}{\nu} \np \conc{\cZ}
\kq{\cZZ}{\cE}
}{
\DD{\cZZ}
}\,,\\
\pcdfluxe{\cE}{\cZ}{\cZZ}{\nu} & =
\frac{
\Big(\dkpe{\cZZ}{\cZ}{\nu} + \dkpse{\cZZ}{\cZ}{\nu}  \,\np\Big)
\kq{\cE}{\cZZ}\conc{\cE}
}{
\DD{\cZZ}
} \,,
\end{align}
\label{eq:flux_pZtoEvZZ}
\end{subequations}
while the corresponding net density current is given by 
\begin{equation}\small
\pcdcurr{\cZ}{\cE}{\cZZ}{\nu}  = 
\pcdfluxa{\cZ}{\cE}{\cZZ}{\nu}
-
\pcdfluxe{\cE}{\cZ}{\cZZ}{\nu}
\,.
\label{eq:currP:Z_ZZ_E}
\end{equation}

The density fluxes of absorption and emission of the cycle in Fig.~\ref{fig:HDM_cycles}c
(or, equivalently, of the \futile\ \reaction\ in Fig~\ref{fig:HDM_effrct}c)
read
\begin{subequations}\small
\begin{align}
\pcdfluxa{\cZ}{\cZ}{\cZZ}{\nu} & = 
\frac{
\dkpa{\cZ}{\cZZ}{\nu} \np \conc{\cZ}
\kq{\cZZ}{\cZ}
}{
\DD{\cZZ}
}\,,\\
\pcdfluxe{\cZ}{\cZ}{\cZZ}{\nu} & =
\frac{
\Big(\dkpe{\cZZ}{\cZ}{\nu} + \dkpse{\cZZ}{\cZ}{\nu}  \,\np\Big)
\kq{\cZ}{\cZZ}\conc{\cZ}
}{
\DD{\cZZ}
} \,,
\end{align}
\label{eq:flux_pZtoZvZZ}
\end{subequations}
while the corresponding net density current is given by 
\begin{equation}\small
\pcdcurr{\cZ}{\cZ}{\cZZ}{\nu}  = 
\pcdfluxa{\cZ}{\cZ}{\cZZ}{\nu}
-
\pcdfluxe{\cZ}{\cZ}{\cZZ}{\nu} 
\,.
\end{equation}

The density flux for the cycle in Fig.~\ref{fig:HDM_cycles}d 
when photons of frequency $\nu$ are absorbed while photons of frequency $\nu'$ are emitted
(or, equivalently, the \futile\ \reaction\ in Fig~\ref{fig:HDM_effrct}d) reads
\begin{equation}\small
\pcddflux{\cZ}{\cZ}{\cZZ}{\nu}{\nu'} = 
\frac{
\dkpa{\cZ}{\cZZ}{\nu} \np \conc{\cZ}
\Big(\dkpe{\cZZ}{\cZ}{\nu'} + \dkpse{\cZZ}{\cZ}{\nu'}  \,\np\Big)
}{
\DD{\cZZ}
}\,,
\label{eq:flux_ppZtoZvZZ}
\end{equation}
while the corresponding net density current is given by 
\begin{equation}\small
\pcddcurr{\cZ}{\cZ}{\cZZ}{\nu}{\nu'} =
\pcddflux{\cZ}{\cZ}{\cZZ}{\nu}{\nu'}
-\pcddflux{\cZ}{\cZ}{\cZZ}{\nu'}{\nu}\,,
\label{eq:curr_ppZtoZvZZ}
\end{equation}
satisfying $ \pcddcurr{\cZ}{\cZ}{\cZZ}{\nu}{\nu'} = - \pcddcurr{\cZ}{\cZ}{\cZZ}{\nu'}{\nu}$.

As long as the time scale separation between excited-state species and ground-state species holds,
i.e., $\conc{\cZZ} = \ssconc{\cZZ}$,
the net cycle currents univocally define the current of all underlying \transitions\ in Fig.~\ref{fig:HDM}.
Mathematically speaking,
the cycle currents are a basis set for the currents of all underlying \transitions .
Indeed, 
the current of the thermally induced \transitions\ can be written as
\begin{subequations}\small
\begin{align}
\qcurr{\cZ}{\cZZ} & =  
-\int\dd\nu\, \pcdcurr{\cZ}{\cZ}{\cZZ}{\nu}  
+ \qccurr{\cZ}{\cE}{\cZZ} 
\,, \label{eq:elcurr_hdm_ZtoZZq}
\\
\qcurr{\cZZ}{\cE}  & = 
\int\dd\nu\, \pcdcurr{\cZ}{\cE}{\cZZ}{\nu}
+ \qccurr{\cZ}{\cE}{\cZZ} 
\,, \label{eq:elcurr_hdm_ZZtoEq}
\end{align}
\label{eq:elcurr_hdm}
\end{subequations}
while the net density current of the photo-induced \transition\ can be written as
\begin{equation}
\begin{split}
\pdcurr{\cZ}{\cZZ}{\nu} = &\,
\pcdcurr{\cZ}{\cZ}{\cZZ}{\nu}
+ \pcdcurr{\cZ}{\cE}{\cZZ}{\nu}
\\
&+\int\dd\nu'\, \pcddcurr{\cZ}{\cZ}{\cZZ}{\nu}{\nu'} 
\,.
\end{split}
\label{eq:elcurr_hdm_ZtoZZp}
\end{equation}


\remark
In order to apply the modified matrix-tree theorem~\cite{Hill1966, HILL197733} used in Ref.~\cite{LiangBusiello2024},
one has to recognize that photo-induced \transitions\ promoted by different frequencies 
have to be treated as independent \transitions .


\subsubsection*{Quantum Yields}

The cycle fluxes in 
Eqs.~\eqref{eq:flux_qZtoEvZZ},~\eqref{eq:flux_pZtoEvZZ},~\eqref{eq:flux_pZtoZvZZ} and~\eqref{eq:flux_ppZtoZvZZ}
can be expressed in terms of the following quantum yields
\begin{subequations}
\begin{align}
\qyq{\cZZ}{\cE} &\equiv  \kq{\cZZ}{\cE} / \DD{\cZZ} \,,\\
\qyq{\cZZ}{\cZ} &\equiv  \kq{\cZZ}{\cZ} / \DD{\cZZ} \,,\\
\qyp{\cZZ}{\cZ}{\nu} &\equiv \big(\dkpe{\cZZ}{\cZ}{\nu} + \dkpse{\cZZ}{\cZ}{\nu}  \,\np\big) / \DD{\cZZ}\,.
\end{align}
\label{eq:QYs}%
\end{subequations}
quantifying the probability that the excited-state species $\cZZ$ interconverts 
to $\cE$ via the thermally induced \transition\ or 
to $\cZ$ via the thermally induced \transition\ or
to $\cZ$ via the photo-induced \transition, 
respectively.
Indeed, by using the definitions~\eqref{eq:QYs} in
Eqs.~\eqref{eq:flux_qZtoEvZZ},~\eqref{eq:flux_pZtoEvZZ},~\eqref{eq:flux_pZtoZvZZ}, and~\eqref{eq:flux_ppZtoZvZZ},
the cycle fluxes become
\begin{subequations}\small
\begin{align}
\qcflux{\cZ}{\cE}{\cZZ} &= \qyq{\cZZ}{\cE} \kq{\cZ}{\cZZ}\conc{\cZ} \,, \\
\qcflux{\cE}{\cZ}{\cZZ} &= \qyq{\cZZ}{\cZ} \kq{\cE}{\cZZ}\conc{\cE} \,, \\
\pcdfluxa{\cZ}{\cE}{\cZZ}{\nu} & = \qyq{\cZZ}{\cE} \dkpa{\cZ}{\cZZ}{\nu} \np \conc{\cZ} \,, \\
\pcdfluxe{\cE}{\cZ}{\cZZ}{\nu} & = \qyp{\cZZ}{\cZ}{\nu} \kq{\cE}{\cZZ}\conc{\cE} \,, \\
\pcdfluxa{\cZ}{\cZ}{\cZZ}{\nu} & = \qyq{\cZZ}{\cZ} \dkpa{\cZ}{\cZZ}{\nu} \np \conc{\cZ} \,, \\
\pcdfluxe{\cZ}{\cZ}{\cZZ}{\nu} & = \qyp{\cZZ}{\cZ}{\nu} \kq{\cZ}{\cZZ}\conc{\cZ} \,, \\
\pcddflux{\cZ}{\cZ}{\cZZ}{\nu}{\nu'} &= \qyp{\cZZ}{\cZ}{\nu'} \dkpa{\cZ}{\cZZ}{\nu} \np \conc{\cZ} \,,
\end{align}
\label{eq:flux_qy}
\end{subequations}
thus proving that
they are functions of the absorption spectra and quantum yields
together with the concentrations of the ground-state species only.

\remark
The quantum yields in Eq.~\eqref{eq:QYs} satisfy
\begin{equation}
\qyq{\cZZ}{\cE}  + \qyq{\cZZ}{\cZ} + \int \dd\nu\,\qyp{\cZZ}{\cZ}{\nu}  = 1
\,.
\end{equation}


\subsection{Rate Equations}

As long as the time scale separation between excited-state species and ground-state species holds,
the concentrations $\conc{\cZ}$ and $\conc{\cE}$ follow
\begin{equation}\small
\dt \conc{\cZ} = -\dt \conc{\cE} =   
- \qccurr{\cZ}{\cE}{\cZZ} 
-\int\dd\nu\, 
\pcdcurr{\cZ}{\cE}{\cZZ}{\nu} 
\,,
\end{equation}
as it can be verified by using Eqs.~\eqref{eq:elcurr_hdm} and~\eqref{eq:elcurr_hdm_ZtoZZp}.
Namely, the dynamics of $\conc{\cZ}$ and $\conc{\cE}$ can be expressed in terms of the net currents
of the cycles (a) and (b) in Fig.~\ref{fig:HDM_cycles}
or equivalently of the \isomerization\ \reactions\  (a) and (b) in Fig.~\ref{fig:HDM_effrct}.

On the other hand,
the density of photons with frequency~$\n{\nu}$ follows
\begin{equation}
\begin{split}
\dt\n{\nu} =& - \bigg[\,
\pcdcurr{\cZ}{\cZ}{\cZZ}{\nu}
+ \pcdcurr{\cZ}{\cE}{\cZZ}{\nu}
\\
&+\int\dd\nu'\, \pcddcurr{\cZ}{\cZ}{\cZZ}{\nu}{\nu'} \bigg] 
+ \excurr
\,.
\end{split}
\end{equation}
as it can be verified by using Eq~\eqref{eq:elcurr_hdm_ZtoZZp}.
Namely, the dynamics of $\n{\nu}$ can be expressed in terms of the net density currents
of the cycles (b), (c), and (d) in Fig.~\ref{fig:HDM_cycles}
or equivalently of the \isomerization\ \reaction\ (b) and the \futile\ \reactions\ (c) and (d) in Fig.~\ref{fig:HDM_effrct}.



\subsection{Local Detailed Balance\label{sub:LDB2}}

By simply plugging Eqs.~\eqref{eq:ldb_qrct_cg1} and~\eqref{eq:ldb_prct_cg1}
into the cycle fluxes in 
Eqs.~\eqref{eq:flux_qZtoEvZZ},~\eqref{eq:flux_pZtoEvZZ},~\eqref{eq:flux_pZtoZvZZ} and~\eqref{eq:flux_ppZtoZvZZ},
we obtain the following local detailed balance conditions for the \reactions\ in Fig.~\ref{fig:HDM_effrct}:
\begin{subequations}\small
\begin{align}
\kb T \ln \frac{
\qcflux{\cZ}{\cE}{\cZZ} 
}{
\qcflux{\cE}{\cZ}{\cZZ} 
} &=
-\big(
\cp{\cE}
- \cp{\cZ}
\big)
\,\\
\kb T \ln \frac{
\pcdfluxa{\cZ}{\cE}{\cZZ}{\nu}
}{
\pcdfluxe{\cE}{\cZ}{\cZZ}{\nu}
} 
&= 
- \big(
\cp{\cE}
- \cp{\cZ}
- \cpp{\nu}
\big)
\,,\\
\kb T \ln \frac{
\pcdfluxa{\cZ}{\cZ}{\cZZ}{\nu}
}{
\pcdfluxe{\cZ}{\cZ}{\cZZ}{\nu}
} 
&= 
\cpp{\nu}
\,,\\
\kb T \ln \frac{
\pcddflux{\cZ}{\cZ}{\cZZ}{\nu}{\nu'} 
}{
\pcddflux{\cZ}{\cZ}{\cZZ}{\nu'}{\nu} 
}
&=
-\big(
\cpp{\nu'} - \cpp{\nu}
\big) 
\,.
\end{align}
\label{eq:ldbHDM}
\end{subequations}

\remark
Like in Subs.~\eqref{sub:LDB1},
if the light source were an actual black body at temperature $\TTbb$,
the chemical potential of the photons~\eqref{eq:cppBB} would be frequency independent and 
local detailed balance condition would then also hold for the fluxes integrated over the frequency, namely,
\begin{subequations}\small
\begin{align}
\kb T 
&\ln \frac{
\int\dd\nu\,
\pcdfluxa{\cZ}{\cE}{\cZZ}{\nu}
}{
\int\dd\nu\,
\pcdfluxe{\cE}{\cZ}{\cZZ}{\nu}
} 
= 
- \big(
\cp{\cE}
- \cp{\cZ}
- \cppbb
\big)
\,,\\
\kb T 
&\ln \frac{
\int\dd\nu\,
\pcdfluxa{\cZ}{\cZ}{\cZZ}{\nu}
}{
\int\dd\nu\,
\pcdfluxe{\cZ}{\cZ}{\cZZ}{\nu}
} 
= 
\cppbb
\,,\\
\kb T 
&\ln \frac{
\int\dd\nu
\int\dd\nu'\,
\pcddflux{\cZ}{\cZ}{\cZZ}{\nu}{\nu'} 
}{
\int\dd\nu
\int\dd\nu'\,
\pcddflux{\cZ}{\cZ}{\cZZ}{\nu'}{\nu} 
}
=
0
\,.
\end{align}
\label{eq:ldbHDM_integrated}
\end{subequations}
with $\cppbb = h \nu (1 - {T}/{\TTbb})$.


\subsection{Second Law\label{sub:SL2}}

We prove here that 
the dynamical description in terms of \reactions\ between ground-state species
is thermodynamically consistent.
Namely, the entropy production rate of the case in Fig.~\ref{fig:HDM}, i.e., 
\begin{equation}
\begin{split}
T\epr 
= 
& -(\cp{\cZZ} - \cp{\cZ}) \qcurr{\cZ}{\cZZ} 
 -(\cp{\cZZ} - \cp{\cE}) \qcurr{\cE}{\cZZ} \\
& -  \int\dd\nu\, ({ \cp{\cZZ} - \cp{\cZ} - \cpp{\nu}}) \pdcurr{\cZ}{\cZZ}{\nu}
\,,
\end{split}
\label{eq:eprHDM_el}
\end{equation}
obtained by specializing Eqs.~\eqref{eq:eprq} and~\eqref{eq:eprp},
can be expressed in terms of the dissipation along the \reactions\ in Fig.~\ref{fig:HDM_effrct} 
(or, equivalently, along the cycles in Fig.~\ref{fig:HDM_cycles}).
Indeed, by plugging Eqs.~\eqref{eq:elcurr_hdm} and~\eqref{eq:elcurr_hdm_ZtoZZp} 
into Eq.~\eqref{eq:eprHDM_el},
we obtain 
\begin{equation}\small
\begin{split}
T\epr 
=
& -(\cp{\cE} - \cp{\cZ}) \qccurr{\cZ}{\cE}{\cZZ}  \\
& -  \int\dd\nu\, (\cp{\cE} - \cp{\cZ} - \cpp{\nu}) \pcdcurr{\cZ}{\cE}{\cZZ}{\nu} \\
& -  \int\dd\nu\, ( -\cpp{\nu}) \pcdcurr{\cZ}{\cZ}{\cZZ}{\nu} \\ 
& -  \frac{1}{2} \int\dd\nu\int\dd\nu'\, (\cpp{\nu'} - \cpp{\nu}) \pcddcurr{\cZ}{\cZ}{\cZZ}{\nu}{\nu'}
\,,
\end{split}
\label{eq:eprHDM}
\end{equation}
where we used $\pcddcurr{\cZ}{\cZ}{\cZZ}{\nu}{\nu'} = - \pcddcurr{\cZ}{\cZ}{\cZZ}{\nu'}{\nu}$.
Each term featuring Eq.~\eqref{eq:eprHDM} is non-negative 
because of the local detailed balance conditions~\eqref{eq:ldbHDM}
and specifies the dissipation of the corresponding \reaction\ in Fig.~\ref{fig:HDM_effrct}.
The first (resp. second) one accounts for the dissipation of the \isomerization\ \reaction\ of $\cZ$ into $\cE$ 
via an underlying sequence of \transitions\ that are 
thermally induced only (resp. both thermally  and photo-induced). 
The third one accounts for the dissipation of the \futile\ \reaction\ of $\cZ$ back to itself
via an underlying sequence of \transitions\ that are both thermally  and photo-induced.
The net effect is the interconversion of photons into heat.
The last one accounts for the dissipation of the \futile\ \reaction\ of $\cZ$ back to itself
via an underlying sequence of \transitions\ that are photo-induced only.
The net effect is the interconversion of photons of frequency $\nu$ into photons of a different frequency $\nu'$.


\section{Coarse Graining of the Diabatic \Process\ \label{sec:DM}}

\begin{figure*}
    \centering
    \includegraphics[width=0.98\textwidth]{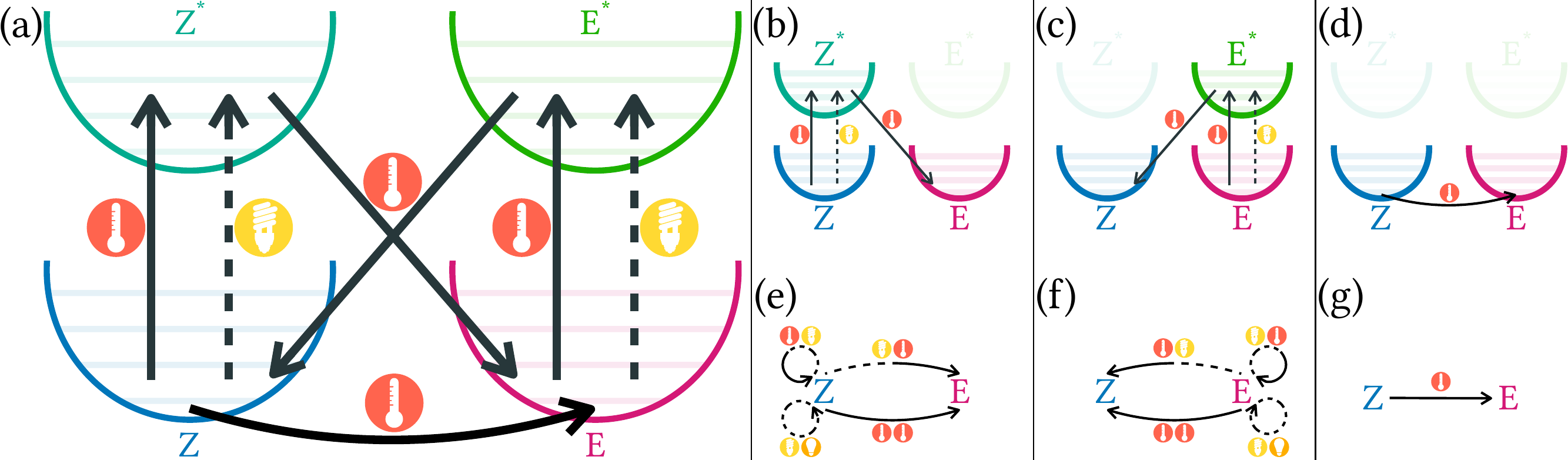}
    \caption{
    Diabatic \process. 
    (a)
    The species in the electronic ground state $\cZ$ (resp. $\cE$) can be interconverted into 
    its electronic excited state $\cZZ$ (resp. $\cEE$)
    via thermally induced (arrow close to a thermometer) 
    and photo-induced (dashed arrow close to a bulb)
    \transitions . 
    The electronic excited state $\cZZ$ (resp. $\cEE$) can also be interconverted 
    via a thermally induced (oblique arrow close to a thermometer) \transition\ 
    into the electronic ground state $\cE$ (resp. $\cZ$) of the other species.
    The ground-state species $\cZ$ 
    can directly interconvert into 
    the ground-state species $\cE$ 
    via a thermally induced transition
    (curved arrow close to a thermometer).
    Chemically speaking, the species $\cZ$ and $\cE$ are different isomers.
    (b) 
    Set of \transitions\ of the diabatic \process\ involving the electronic excited-state species $\cZZ$.
    When the dynamics of $\cZZ$ is faster compared to $\cZ$ and $\cE$, 
    these \transitions\ can be coarse grained into the (\isomerization\ and \futile ) \reactions\ in (e) between $\cZ$ and $\cE$ 
    involving the excited-state species $\cZZ$.
    (c) 
    Set of \transitions\ of the diabatic \process\ involving the electronic excited-state species $\cEE$.
    When the dynamics of $\cEE$ is faster compared to $\cZ$ and $\cE$, 
    these \transitions\ can be coarse grained into the (\isomerization\ and \futile ) \reactions\ in (f) between $\cZ$ and $\cE$
    involving the excited-state species $\cEE$.
    (d)
    \Transition\ that does not involve electronic excited-state species $\cZZ$ and $\cEE$
    and corresponds to the direct \isomerization\ \reaction\ of $\cZ$ into $\cE$ in (g).
    \rev{Note that each \transition\ and \reaction\ has always a backward counterpart, even if it is not represented.}
    \label{fig:DM}}
\end{figure*}

We now use the results of Sec.~\ref{sec:HDM} to coarse grain the photo-isomerization scheme in Fig.~\ref{fig:DM}a,
known as diabatic \process~\cite{foerster1970,Balzani2014}, in terms of \reactions\ between ground-state species only.
The diabatic \process\ describes 
the interconversion of one species $\cZ$ (in the electronic ground state)
into a different isomer $\cE$ (in the electronic ground state) 
which involves the electronic excited-state species $\cZZ$ and $\cEE$.
It is widely common in organic molecules that undergo photo-isomerizations involving,
for example, alkene~\cite{cerullo2010} or azobenzene~\cite{mukamel2021} derivatives.

We start by recognizing that the \transitions\ in Fig.~\ref{fig:DM}a can be grouped into three sets 
illustrated in Fig.~\ref{fig:DM}b-c-d, respectively.
The first set (Fig.~\ref{fig:DM}b) collects 
the same \transitions\ discussed in Sec.~\ref{sec:HDM},
namely, the thermally  and photo-induced \transitions\
involving the excited-state species $\cZZ$ and leading to the interconversion of $\cZ$ into $\cE$.
The second set (Fig.~\ref{fig:DM}c) collects 
the thermally  and photo-induced \transitions\
involving the excited-state species $\cEE$ and leading to the interconversion of $\cE$ into $\cZ$.
In practice, it collects the same \transitions\ discussed in Sec.~\ref{sec:HDM}, 
after relabelling the species according to $\cZ$ into $\cE$, $\cE$ into $\cZ$, and $\cEE$ into $\cZZ$.
The third set (Fig.~\ref{fig:DM}d) collects the direct interconversion of $\cZ$ into $\cE$ 
via a thermally induced \transition , 
which does not involve any excited-state species.

The first two sets of \transitions\ (in Fig.~\ref{fig:DM}b and Fig.~\ref{fig:DM}c, respectively)
can be coarse grained by repeating the procedure illustrated in Sec.~\ref{sec:HDM}.
By assuming that the evolution of the excited-state species $\cZZ$ and $\cEE$ is much faster than 
the evolution of the ground-state species $\cZ$ and $\cE$,
the latter can be described in terms of cycles 
defining (\isomerization\ and \futile ) \reactions\ between the ground-state species $\cZ$ and $\cE$ only.
The \reactions\ for the first set (Fig.~\ref{fig:DM}e), 
as well as their fluxes and net currents, are those derived in Sec.~\ref{sec:HDM}.
Similarly, the \reactions\ for the second set (Fig.~\ref{fig:DM}f), 
as well as their fluxes and net currents, correspond to those derived in Sec.~\ref{sec:HDM}
after relabelling the species according to $\cZ$ into $\cE$, $\cE$ into $\cZ$, and $\cEE$ into $\cZZ$.
Crucially, all these \reactions\ still satisfy a local detailed balance condition
as proved in  Sec.~\ref{sec:HDM}.
The third and last set of \reactions\ is the direct interconversion of $\cZ$ into $\cE$
and satisfies a local detailed balance condition as discussed in Sec.~\ref{sec:cg_vibrational_states}.

Therefore,
as long as the time scale separation 
between excited-state species and ground-state species holds, 
the concentrations $\conc{\cZ}$ and $\conc{\cE}$ follow
\begin{equation}\small
\begin{split}
\dt \conc{\cZ} = -\dt \conc{\cE}
=  
& - \qccurr{\cZ}{\cE}{\cZZ} - \int\dd\nu\, \pcdcurr{\cZ}{\cE}{\cZZ}{\nu} \\
&+ \qccurr{\cE}{\cZ}{\cEE} +\int\dd\nu\, \pcdcurr{\cE}{\cZ}{\cEE}{\nu} \\
&- \qcurr{\cZ}{\cE}
\,,
\end{split}
\label{eq:req_DM}
\end{equation}
where 
the first,
second, 
and third 
line feature the currents 
(Eqs.~\eqref{eq:currQ:Z_ZZ_E}, \eqref{eq:currP:Z_ZZ_E}, and~\eqref{eq:cg1currq}) 
of the isomeization reactions involving 
$\cZZ$ (Fig.~\ref{fig:DM}e), 
$\cEE$ (Fig.~\ref{fig:DM}f), 
and no excited-state species (Fig.~\ref{fig:DM}g), 
respectively.
However, 
the fluxes of the isomerization \reactions\
involving thermally induced \transitions\
from a ground-state species to an excited-state species (see Eq.~\eqref{eq:flux_qy})
are usually negligible compared to the others.
The rate equation~\eqref{eq:req_DM} can thus be simplified to 
\begin{equation}\small
\begin{split}
\dt \conc{\cZ} = -\dt \conc{\cE}
\approx
&-\int\dd\nu\, \qyq{\cZZ}{\cE} \dkpa{\cZ}{\cZZ}{\nu} \np \conc{\cZ}\\
&+\int\dd\nu\, \qyq{\cEE}{\cZ} \dkpa{\cE}{\cEE}{\nu} \np \conc{\cE} \\
&- \bigg( \kq{\cZ}{\cE}\conc{\cZ} - \kq{\cE}{\cZ}\conc{\cE}\bigg)
\,,
\end{split}
\label{eq:req_DM_simplified}
\end{equation}
by expressing the reaction fluxes according to
Eqs.~\eqref{eq:flux_q_e} and~\eqref{eq:flux_qy}.
We notice that Eq.~\eqref{eq:req_DM_simplified} corresponds to the stardard rate equation 
used to describe the dynamics of the diabatic \process\
in terms of the experimentally measurable 
kinetic constants, 
absorption spectra,
and quantum yields~\cite{Montalti2006}.
By factorizing the concentrations $\conc{\cZ}$ and $\conc{\cE}$,
Eq.~\eqref{eq:req_DM_simplified} is often interpreted as the rate equation 
resulting from the two effective reactions
$\cZ \ch{->} \cE$
and 
$\cE \ch{->} \cZ$,
whose kinetic constants read
$\big[\int\dd\nu\, \qyq{\cZZ}{\cE} \dkpa{\cZ}{\cZZ}{\nu} \np + \kq{\cZ}{\cE}\big]$
and
$\big[\int\dd\nu\, \qyq{\cEE}{\cZ} \dkpa{\cE}{\cEE}{\nu} \np + \kq{\cE}{\cZ}\big]$, 
respectively.


\section{Implications of the Local Detailed Balance for Photo-Driven Molecular \Ratchets\label{sec:MM}}

\begin{figure*}
    \centering
    \includegraphics[width=0.98\textwidth]{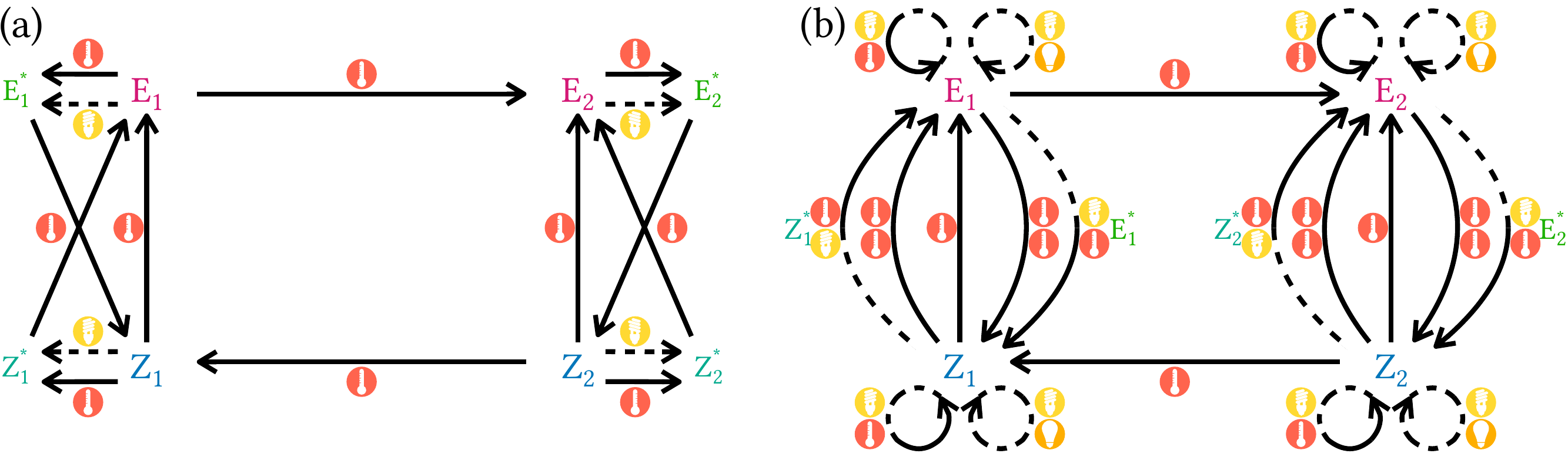}
    \caption{
    \rev{Mechanism of prototypical autonomous photo-driven molecular \ratchets .
    (a) 
    Network of \transitions\ explicitly accounting for the excited-state species
    $\cZZa$, $\cEEa$, $\cZZb$ and $\cEEb$.
    The two horizontal long arrows (close to a thermometer) represent the thermally induced \reactions\ interconverting 
    $\cZa$ into $\cZb$, on the one hand, and $\cEa$ into $\cEb$, on the other hand.
    The other arrows represent the diabatic \processes\ 
    interconverting 
    $\cZa$ into $\cEa$
    (via \transitions\ also involving the excited-state species $\cZZa$ and $\cEEa$), on the one hand, 
    and $\cZb$ into $\cEb$
    (via \transitions\ also involving the excited-state species $\cZZb$ and $\cEEb$), on the other hand.
    (b)
    Network of \reactions\ between ground-state species 
    resulting from the coarse graining of the diabatic processes 
    (discussed in Sec.~\ref{sec:DM} and illustrated in Fig.~\ref{fig:DM}).
    The horizontal arrows (close to a thermometer) still represent the thermally induced \reactions\ interconverting 
    $\cZa$ into $\cZb$, on the one hand, and $\cEa$ into $\cEb$, on the other hand.
    The other arrows represent the (\isomerization\ and \futile ) \reactions\ (given in Fig.~\ref{fig:DM}e-f-g)
    resulting from the coarse graining of the diabatic \processes\ 
    interconverting $\cZa$ into $\cEa$, on the one hand, and $\cZb$ into $\cEb$, on the other hand.
    Note that all \transitions\ and \reactions\ have always their backward counterparts, even if they are not represented.}
    \label{fig:MM}}
\end{figure*}

We now consider a prototypical autonomous photo-driven molecular \ratchet ,
whose network of \rev{\transitions } 
is illustrated in Fig.~\ref{fig:MM}\rev{a}.
On the one hand,
thermally induced \reactions\
interconvert the (ground-state) species $\cZa$ into  $\cZb$ and  $\cEa$ into $\cEb$.
On the other hand, 
diabatic \processes\ (discussed in Sec.~\ref{sec:DM} and illustrated in Fig.~\ref{fig:DM})
interconvert the (ground-state) species $\cZa$ into $\cEa$ 
\rev{(via \transitions\ also involving the excited-state species $\cZZa$ and $\cEEa$)}
and 
$\cZb$ into $\cEb$
\rev{(via \transitions\ also involving the excited-state species $\cZZb$ and $\cEEb$)}.
This scheme is routinely 
\rev{coarse grained in terms of reactions between ground-state species only
and}
used to exemplify 
the conceptual and practical elements of photo-driven molecular \ratchets~\cite{Ragazzon2023review, Borsley2024,astumian2024review}.
\rev{Its validity has been verified} against experimentally realized molecular 
motors~\cite{pooler2021} and pumps~\cite{leigh2007,Corra2022a}.
\rev{The photo-isomerization and futile reactions resulting 
from the (thermodynamically consistent)
coarse graining (developed in Sec.~\ref{sec:DM}) of the diabatic processes in Fig.~\ref{fig:MM}a
are illustrate in Fig.~\ref{fig:MM}b.}


We examine the implications that the existence of a local detailed balance condition
for \rev{(photo-isomerization and futile reactions of) the} diabatic \processes\ has on 
the kinetic asymmetry of the network~\cite{astumian2019},
quantified by the ratcheting constant $\Kr$~\cite{Ragazzon2023review,Borsley2024, corra2025review}.
This constant serves as a directionality parameter,
identifying whether, at steady state, 
the net currents across
the network of \reactions\ in Fig.~\ref{fig:MM}\rev{b}
are preferentially directed
clockwise (corresponding to $\Kr > 1$),
counterclockwise (corresponding to $\Kr < 1$),
or show no net bias (corresponding to $\Kr = 1$).
For the network of \reactions\ in Fig.~\ref{fig:MM}\rev{b}, 
$\Kr$ can be written as
\begin{widetext}
\begin{equation}
\begin{split}
\Kr = &\frac{
\bigg(
\bigintssss\dd\nu\, \pcdfluxa{\cZa}{\cEa}{\cZZa}{\nu} + 
\qcflux{\cZa}{\cEa}{\cZZa} + 
\qflux{\cZa}{\cEa} + 
\qcflux{\cZa}{\cEa}{\cEEa}+
\bigintssss\dd\nu\, \pcdfluxe{\cZa}{\cEa}{\cEEa}{\nu}
\bigg)
\qflux{\cEa}{\cEb} 
}{
\bigg(
\bigintssss\dd\nu\, \pcdfluxe{\cEa}{\cZa}{\cZZa}{\nu} + 
\qcflux{\cEa}{\cZa}{\cZZa} + 
\qflux{\cEa}{\cZa} + 
\qcflux{\cEa}{\cZa}{\cEEa}+
\bigintssss\dd\nu\, \pcdfluxa{\cEa}{\cZa}{\cEEa}{\nu}
\bigg)
\qflux{\cEb}{\cEa} 
}\times \\
&\frac{
\bigg(
\bigintssss\dd\nu\, \pcdfluxe{\cEb}{\cZb}{\cZZb}{\nu} + 
\qcflux{\cEb}{\cZb}{\cZZb} + 
\qflux{\cEb}{\cZb} + 
\qcflux{\cEb}{\cZb}{\cEEb}+
\bigintssss\dd\nu\, \pcdfluxa{\cEb}{\cZb}{\cEEb}{\nu}
\bigg)
\qflux{\cZb}{\cZa} 
}{
\bigg(
\bigintssss\dd\nu\, \pcdfluxa{\cZb}{\cEb}{\cZZb}{\nu} + 
\qcflux{\cZb}{\cEb}{\cZZb} + 
\qflux{\cZb}{\cEb} + 
\qcflux{\cZb}{\cEb}{\cEEb}+
\bigintssss\dd\nu\, \pcdfluxe{\cZb}{\cEb}{\cEEb}{\nu}
\bigg)
\qflux{\cZa}{\cZb}
}\,,
\end{split}
\label{eq:kr_def}
\end{equation}
\end{widetext}
where the four terms in parentheses are
the net fluxes resulting from
the sum of the fluxes of all \reactions\ leading to the interconversions
$\cZa \to \cEa$,
$\cEa \to \cZa$,
$\cEb \to \cZb$,
and $\cZb \to \cEb$,
respectively.
Namely,
the numerator features the product of the net fluxes of the \reactions\ cycling clockwise,
i.e., $\cZa \to \cEa \to \cEb \to \cZb \to \cZa$,
while the denominator features the product of the net fluxes of the \reactions\ cycling counterclockwise,
i.e., $\cZa \gets \cEa \gets \cEb \gets \cZb \gets \cZa$.

We first stress that 
the expression of the ratcheting constant
only resembles 
\----
but is not
\----
that of (the exponential of) a thermodynamic affinity.
Indeed,
only if each net flux (in parentheses) in Eq.~\eqref{eq:kr_def} 
were the flux of a single reaction
satisfying local detailed balance
\----
which is not the case for the photo-driven molecular \ratchet\ in Fig.~\ref{fig:MM}\rev{b}
\----
then 
would $\kb T \ln \Kr$ correspond to
the thermodynamic cycle affinity.

We then notice that the expression of the ratcheting constant in Eq.~\eqref{eq:kr_def} 
might seem different from the usual one~\cite{Ragazzon2023review,corra2025review}.
On the one hand, 
$\Kr$ is usually written in terms of kinetic constants rather than \reaction\ fluxes,
but the two expressions are equivalent.
Indeed, by using the fluxes in 
Eqs.~\eqref{eq:flux_q_e} and~\eqref{eq:flux_qy},
the dependence on the concentrations of the ground-state species,
i.e., $\conc{\cZa}$, $\conc{\cZb}$, $\conc{\cEa}$, and $\conc{\cEb}$,
cancels out 
and the ratcheting constant becomes a function of kinetic constants only 
or, equivalently, 
the absorption spectra and quantum yields 
(see Eqs.~\eqref{eq:flux_q_e} and~\eqref{eq:flux_qy}).
We chose not to write $\Kr$ in Eq.~\eqref{eq:kr_def} 
(as well as in Eq.~\eqref{eq:Kr_ldb})
in terms of kinetic constants
to maintain a more compact notation.
On the other hand, 
the terms in parentheses
accounting for to the interconversions
$\cZa \to \cEa$,
$\cEa \to \cZa$,
$\cEb \to \cZb$,
and $\cZb \to \cEb$,
are usually written as the sum of few fluxes~\cite{ragazzon2018, Ragazzon2023review, corra2025review},
while each of these terms in Eq.~\eqref{eq:kr_def} is given by the sum of five fluxes.
The reason for this difference arises from the coarse graining of the diabatic \process.
As explained in Subs.~\ref{sec:cghdm:kinetics},
we identified the (\isomerization\ and \futile ) \reactions\ that still satisfy the local detailed balance condition
by constructing them using all possible cycles between ground-state species.
This led to five \reactions .
Nevertheless, the usual expression of $\Kr$ can be recovered from  Eq.~\eqref{eq:kr_def}
by
i) neglecting the (usually small) fluxes of the isomerization \reactions\
involving thermally induced \transitions\
from a ground-state species to an excited-state species (see Eq.~\eqref{eq:flux_qy})
and 
ii) simplifying the concentration dependences using Eqs.~\eqref{eq:flux_q_e} and~\eqref{eq:flux_qy}.
Indeed, this leads to
\begin{equation}\small
\begin{split}
\Kr 
\approx 
&\frac{
\bigg(
\bigintssss\dd\nu\, \qyq{\cZZa}{\cEa} \dkpa{\cZa}{\cZZa}{\nu} \np  + 
\kq{\cZa}{\cEa}
\bigg)
\kq{\cEa}{\cEb}
}{
\bigg(
\kq{\cEa}{\cZa} +
\bigintssss\dd\nu\, \qyq{\cEEa}{\cZa} \dkpa{\cEa}{\cEEa}{\nu} \np 
\bigg)
\kq{\cEb}{\cEa}
}\times \\
&\frac{
\bigg(
\kq{\cEb}{\cZb} +
\bigintssss\dd\nu\, \qyq{\cEEb}{\cZb} \dkpa{\cEb}{\cEEb}{\nu} \np 
\bigg)
\kq{\cZb}{\cZa}
}{
\bigg(
\bigintssss\dd\nu\, \qyq{\cZZb}{\cEb} \dkpa{\cZb}{\cZZb}{\nu} \np  +
\kq{\cZb}{\cEb}
\bigg)
\kq{\cZa}{\cZb}
}\,.
\end{split}
\label{eq:Kr_simplified}
\end{equation}
While Eq.~\eqref{eq:kr_def} and Eq.~\eqref{eq:Kr_simplified}
(approximately) provide the same numerical value of the ratcheting constant,
only the former allows us to identify 
crucial properties of the working mechanism of photo-driven molecular ratchets.

To do so, 
we now use the local detailed balance conditions 
in Eqs.~\eqref{eq:ldb_qrct_cg1} and~\eqref{eq:ldbHDM} 
in the ratcheting constant in Eq.~\eqref{eq:kr_def},
thus obtaining 
\begin{widetext}
\begin{equation}
\begin{split}
\Kr = &\frac{
\bigg(
\bigintssss\dd\nu\, \pcdfluxa{\cZa}{\cEa}{\cZZa}{\nu} + 
\qcflux{\cZa}{\cEa}{\cZZa} + 
\qflux{\cZa}{\cEa} + 
\qcflux{\cZa}{\cEa}{\cEEa}+
\bigintssss\dd\nu\, \pcdfluxe{\cZa}{\cEa}{\cEEa}{\nu}
\bigg)
}{
\bigg(
\bigintssss\dd\nu\, \pcdfluxa{\cZa}{\cEa}{\cZZa}{\nu} e^{-\frac{\cpp{\nu}}{\kb T}}+ 
\qcflux{\cZa}{\cEa}{\cZZa} + 
\qflux{\cZa}{\cEa} + 
\qcflux{\cZa}{\cEa}{\cEEa}+
\bigintssss\dd\nu\, \pcdfluxe{\cZa}{\cEa}{\cEEa}{\nu} e^{\frac{\cpp{\nu}}{\kb T}}
\bigg)
}\times \\
&\frac{
\bigg(
\bigintssss\dd\nu\, \pcdfluxa{\cZb}{\cEb}{\cZZb}{\nu} e^{-\frac{\cpp{\nu}}{\kb T}} + 
\qcflux{\cZb}{\cEb}{\cZZb} + 
\qflux{\cZb}{\cEb} + 
\qcflux{\cZb}{\cEb}{\cEEb}+
\bigintssss\dd\nu\, \pcdfluxe{\cZb}{\cEb}{\cEEb}{\nu} e^{\frac{\cpp{\nu}}{\kb T}}
\bigg)
}{
\bigg(
\bigintssss\dd\nu\, \pcdfluxa{\cZb}{\cEb}{\cZZb}{\nu} + 
\qcflux{\cZb}{\cEb}{\cZZb} + 
\qflux{\cZb}{\cEb} + 
\qcflux{\cZb}{\cEb}{\cEEb}+
\bigintssss\dd\nu\, \pcdfluxe{\cZb}{\cEb}{\cEEb}{\nu}
\bigg)
}\,,
\end{split}
\label{eq:Kr_ldb}
\end{equation}
\end{widetext}
which is independent of the concentrations of the ground-state species,
i.e., $\conc{\cZa}$, $\conc{\cZb}$, $\conc{\cEa}$, and $\conc{\cEb}$
even if it is written in terms of reaction fluxes
(see Eqs.~\eqref{eq:flux_q_e} and~\eqref{eq:flux_qy}).

Three important properties of the working mechanism of 
photo-driven \ratchets\ follow from Eq.~\eqref{eq:Kr_ldb}.
\rev{The first one is well known, 
while the other two result from the fact that 
the isomerization reactions satisfy the local detailed balance condition.
We now examine each of these properties
and explain their physical meaning.}
\rev{First,} 
if $\cpp{\nu} = 0$,
then $\Kr = 1$ for any value of the \reaction\ fluxes and directionality cannot emerge. 
This is a direct implication of fact that the radiation is in equilibrium with the solution when $\cpp{\nu} = 0$.
Hence,
no (nonconservative~\cite{Rao2016}) themodynamic forces maintain \reactions\ out of equilibrium
and the steady state is an equilibrium state satisfying detailed balance.
\rev{Second},
the fluxes of the thermally induced \reactions\ 
$\cZa \ch{<=>} \cZb$ 
and
$\cEa \ch{<=>} \cEb$,
i.e., $\qflux{\cEa}{\cEb}$, $\qflux{\cEb}{\cEa}$, $\qflux{\cZb}{\cZa}$, and $\qflux{\cZa}{\cZb}$,
do not enter the expression of $\Kr$ in Eq.~\eqref{eq:Kr_ldb}.
This implies that the equilibrium constants of these \reactions , i.e.,
\begin{subequations}
\begin{align}
\kqeq{\cEa}{\cEb}  & =\frac{\kq{\cEa}{\cEb}}{\kq{\cEb}{\cEa}}\,, \\
\kqeq{\cZa}{\cZb}  & =\frac{\kq{\cZa}{\cZb}}{\kq{\cZb}{\cZa}}\,,
\end{align}
\label{eq:Keq_rm}
\end{subequations}
do not play any role in determining the directionality.
\rev{Third,}
if all the kinetic constants 
(or, equivalently, 
the absorption spectra and quantum yields) 
of the \reactions\ interconverting $\cZa$ into $\cEa$ 
were equal to 
those of the \reactions\ interconverting $\cZb$ into $\cEb$,
namely,
\begin{subequations}
\begin{align}
\pcdfluxa{\cZa}{\cEa}{\cZZa}{\nu}/\conc{\cZa} 
& = \pcdfluxa{\cZb}{\cEb}{\cZZb}{\nu}/\conc{\cZb} \,, \\
\qcflux{\cZa}{\cEa}{\cZZa}/\conc{\cZa}
& = \qcflux{\cZb}{\cEb}{\cZZb}/\conc{\cZb} \,,\label{eq:no_info_ratchet_m1}\\
\qflux{\cZa}{\cEa}/\conc{\cZa} 
& = \qflux{\cZb}{\cEb}/\conc{\cZb} \,, \\
\qcflux{\cZa}{\cEa}{\cEEa} /\conc{\cZa}
& = \qcflux{\cZb}{\cEb}{\cEEb}/\conc{\cZb} \,,\label{eq:no_info_ratchet_m2}\\
\pcdfluxe{\cZa}{\cEa}{\cEEa}{\nu}/\conc{\cZa}
& = \pcdfluxe{\cZb}{\cEb}{\cEEb}{\nu}/\conc{\cZb} \,,
\end{align}\label{eq:no_info_ratchet}
\end{subequations}
then $\Kr = 1$ and directionality could not emerge.
We conclude by rewritting Eq.~\eqref{eq:Kr_ldb}
in a more standard form.
To do so,
we 
i) 
neglect the (usually small) fluxes of the thermally induced isomerization reactions 
(involving thermally induced transitions 
from a ground-state species to an excited-state species)
and 
ii)
express the remaining fluxes using Eqs.~\eqref{eq:flux_q_e} and~\eqref{eq:flux_qy}.
We thus obtain
\begin{widetext}
\begin{equation}
\Kr 
\approx
\frac{
\left[
\frac{
\bigintssss\dd\nu\, \qyq{\cZZa}{\cEa} \dkpa{\cZa}{\cZZa}{\nu} \np 
+ \bigintssss\dd\nu\, \qyp{\cEEa}{\cEa}{\nu} \kq{\cZa}{\cEEa}
}{
\kq{\cZa}{\cEa}
} 
+ 1
\right]
\left[
\frac{ 
\bigintssss\dd\nu\,  \qyq{\cZZb}{\cEb} \dkpa{\cZb}{\cZZb}{\nu} \np e^{-\frac{\cpp{\nu}}{\kb T}}
+ \bigintssss\dd\nu\, \qyp{\cEEb}{\cEb}{\nu} \kq{\cZb}{\cEEb} e^{\frac{\cpp{\nu}}{\kb T}}
}{
\kq{\cZb}{\cEb}
}
+ 1
\right]
}{
\left[ 
\frac{\bigintssss\dd\nu\,\qyq{\cZZa}{\cEa} \dkpa{\cZa}{\cZZa}{\nu} \np e^{-\frac{\cpp{\nu}}{\kb T}}
+ \bigintssss\dd\nu\, \qyp{\cEEa}{\cEa}{\nu} \kq{\cZa}{\cEEa} e^{\frac{\cpp{\nu}}{\kb T}}
}{
\kq{\cZa}{\cEa}
}
+1
\right]
\left[
\frac{\bigintssss\dd\nu\, \qyq{\cZZb}{\cEb} \dkpa{\cZb}{\cZZb}{\nu} \np 
+ \bigintssss\dd\nu\, \qyp{\cEEb}{\cEb}{\nu} \kq{\cZb}{\cEEb}
}{
\kq{\cZb}{\cEb}
}
+ 1
\right]
}
\,.
\label{eq:Kr_ldb_simplified}
\end{equation}
\end{widetext} 
Equation~\eqref{eq:Kr_ldb_simplified}
still shows that
directionality is not affected by the equilibrium constants in Eq.~\eqref{eq:Keq_rm}
and does not emerge if $\cpp{\nu} = 0$.
It also explicitly shows that 
the following equivalence between
(experimentally measurable)
kinetic constants and quantum yields,
\begin{subequations}
\begin{align}
\qyq{\cZZa}{\cEa}\dkpa{\cZa}{\cZZa}{\nu} &= \qyq{\cZZb}{\cEb}\dkpa{\cZb}{\cZZb}{\nu} \,,
\end{align}
\end{subequations}
is not a sufficient condition to get $\Kr = 1$.


\subsection{Ratchet Mechanism}

We now discuss the meaning of  
the properties of photo-driven \ratchets\ derived from Eq.~\eqref{eq:Kr_ldb}
in terms of ratchet mechanisms.

To do so, 
we start by recalling that
molecular \ratchets\ primarily operate through two distinct  
(but not mutually exclusive) mechanisms:
the energy ratchet and the information ratchet mechanisms~\cite{Ragazzon2023review,Borsley2024,astumian2024review,corra2025review}.
In the former, 
directionality is dictated by differences in the free energy of reaction intermediates 
\rev{or, equivalently,} equilibrium constants~\cite{penocchio2024chem}.
\rev{In} the latter, 
directionality is dictated by differences in the fluxes of the free-energy-harnessing reactions\rev{, e.g., the vertical reactions in Fig.~\ref{fig:MM}b,}
determined by the information encoded in the intermediates they interconvert~\cite{ragazzon2018}.
The independence of the ratcheting constant~\eqref{eq:Kr_ldb} from the equilibrium constants~\eqref{eq:Keq_rm}
shows that the mechanism underlying directionality in autonomous photo-driven \ratchets\
does not rely on the energy differences between reaction intermediates.
Namely, there is no energy ratchet contribution.
The fact that the kinetic constants 
of the diabatic \processes\ 
interconverting $\cZa$ into $\cEa$ and $\cZb$ into $\cEb$
must differ (i.e., Eq.~\eqref{eq:no_info_ratchet} must not hold)
to ensure $\Kr \neq 1$
means that the mechanism underlying directionality in autonomous photo-driven \ratchets\
relies on differences in the fluxes of the free-energy-harnessing reactions.
Namely, the ratchet mechanism is a pure information ratchet mechanism.

Until now, 
autonomous chemically driven and photo-driven \ratchets\ 
were considered fundamentally different because only the former were believed 
to operate via a pure information ratchet mechanism.
Indeed, so far, 
only thermally induced \reactions\ underpinning chemically driven \ratchets\ were known to
satisfy the local detailed balance condition,
implying that
differences in the free energy of reaction intermediates are 
irrelevant to determine directionality~\cite{astumian2015irrelevance,astumian2019}
(see also App.~\ref{app:CDratchets}).
At the same time,
(approximated) expressions like the one in Eq.~\eqref{eq:Kr_simplified} 
led to the (erroneous) idea that 
the ratchet mechanism of autonomous photo-driven \ratchets\ might include an energy ratchet contribution~\cite{astumian2016,Ragazzon2023review,Borsley2024,astumian2024review,corra2025review}.
By establishing a local detailed balance condition also for the (\isomerization\ and \futile ) \reactions\ of the diabatic \process,
we proved that this is not the case:
the ratchet mechanism of 
autonomous photo-driven \ratchets\ is exactly the same as that of chemically driven ones,
i.e., an information ratchet mechanism.

\rev{
This result reveals a paradigmatic shift in the design and optimization principles of photo-driven ratchets
that can be now directly built on top of the insights already well established for chemically driven ratchets.
Specifically,
rather than focusing on increasing energy differences between reaction intermediates, 
design and optimization strategies should aim to maximize the differentiation between the reaction fluxes 
of the free-energy-harnessing reactions,
as this is the primary factor affecting the ratcheting constant~\eqref{eq:Kr_ldb} and,
therefore, directionality. }

\rev{Finally, we notice that 
the conclusion that autonomous photo-driven molecular \ratchets\
operate exclusively through an information ratchet mechanism
is derived from the network of reactions in Fig.~\ref{fig:MM}\rev{b}.
This does not diminish its relevance, since
i)
many autonomous photo-driven molecular ratchets are in fact well described by this network~\cite{pooler2021, leigh2007, Corra2022a}
and 
ii) the design and optimization principles inferred from this network
have been shown to remain valid for more complex cases~\cite{ragazzon2025activetransport, penocchio2024chem}.
Furthermore, 
the definition of the ratcheting constant $K_r$ in Eq.~\eqref{eq:kr_def} extends naturally to multi-cycle photochemical networks 
as well~\cite{Yang2025},
and so does our conclusion.
That said, 
we acknowledge that there also exist photo-driven molecular ratchets whose underlying mechanism 
does not involve photo-isomerizations~\cite{Lehn2006}.
In such cases, our conclusion does not apply, at least not in its present formulation.}



\section{A brief history of Local Detailed Balance\label{sec:micro_rev}}

Before concluding, 
we offer a short historical perspective on the local detailed balance condition to contextualize our work.

The local detailed balance condition is a constraint 
on the ratio of 
the fluxes of each pair of forward and
backward elementary transitions
between suitably coarse-grained states describing a system.
It guarantees the thermodynamic consistency of the description 
and, in the absence of (nonconservative) thermodynamic forces, 
the relaxation of the system to equilibrium.
This constraint has emerged independently in different contexts 
and has been referred to by various names across disciplines.
Indeed, 
it is often called ``microscopic reversibility''~\cite{Blackmond2009,Astumian:2012aa} 
or ``generalized detailed balance''~\cite{Peliti2021}.
Since systems undergoing thermally and photo-induced transitions/reactions
provide two early contexts
where the local detailed balance was introduced, 
we take this opportunity 
to briefly revisit the historical development of this constraint.

To the best of our knowledge, 
the earliest instance of a local detailed balance condition is due to Boltzmann, 
who introduced it in the context of the Boltzmann equation 
describing the dynamics of dilute gases~\cite{Tolman1938}.
There, Boltzmann assumed a fundamental time-reversal symmetry between 
the rates describing the change in momentum caused by molecular collisions that 
ensures the validity of the H-theorem and thus relaxation towards equilibrium~\cite{Forastiere2024}.
Boltzmann’s original work was formulated within the microcanonical ensemble.
Later generalizations to the canonical ensemble, 
relevant for thermally or photo-induced transitions/reactions, 
emerged independently across multiple fields.

For systems undergoing (thermally induced) chemical reactions,
Marcelin formulated an early version of the local detailed balance condition
in Eq.~\eqref{eq:ldb_qrct}.
He expressed this
in terms of what he introduced as 
the standard Gibbs free energy of activation,
namely, 
the free energy of the common activated state 
that the pathways of both the forward and backward 
(thermally induced) chemical reactions
must overcome~\cite{marcelin1915}.
For this reason, 
Marcelin’s work is also regarded as 
the first theoretical treatment 
that justifies the Arrhenius equation from first principles,
by laying the groundwork for the transition state theory~\cite{laidler1985}.
Following similar approaches, other researchers, 
most notably Tolman~\cite{tolman1925} and Lewis~\cite{lewis1925microrev},
introduced the principle of microscopic reversibility 
for (thermally induced) chemical reactions, 
leading to the same constraint.
It is worth noting that in their original formulations~\cite{tolman1925,lewis1925microrev},
the principle of microscopic reversibility was introduced 
as a dynamic equilibrium condition,
namely, that the net current of each elementary transition/reaction vanishes at equilibrium,
rather than in the form of Eq.~\eqref{eq:ldb_qrct}, 
although the latter follows directly from the former.
Subsequently, Van Rysselberghe proposed the first general framework for 
incorporating thermodynamic constraints in networks of 
(thermally induced) elementary transitions/reactions 
arbitrarily far from equilibrium~\cite{VanRysselberghe1958}.

For systems undergoing photo-induced transitions/reactions, 
as emphasized in the main text, 
Einstein’s thermodynamic argument for postulating the existence of the spontaneous emission 
can be considered as another early derivation of 
the local detailed balance condition.
Indeed, 
although perturbation theory of an atom in a classical radiation field predicts only
absorption and stimulated emission, 
Einstein argued that thermodynamic consistency requires 
an additional emission process~\cite{einstein1917,Cohen-Tannoudji1998}.
By requiring that an atom immersed in a radiation field reaches thermal equilibrium 
with the blackbody spectrum, 
Einstein derived 
that the ratio of 
the fluxes of absorption and total (spontaneous and stimulated) emission are governed by a Boltzmann factor, 
as expressed in Eq.~\eqref{eq:ldb_prct}.
Notably, in quantum electrodynamics, 
both spontaneous and stimulated emission arise from the same elementary interaction between
matter and the quantized electromagnetic field.
From this modern perspective, 
spontaneous and stimulated emission are not fundamentally distinct processes,
namely, they represent different limits of the same physical mechanism~\cite{Cohen-Tannoudji1997}.
This realization remains underappreciated in most of the chemistry community, 
where photo-induced transitions are often treated as 
if they were unconstrained by microscopic reversibility~\cite{iupac2009}.

Finally, the terminology ``local detailed balance'' itself was introduced in 
a 1983 paper by Katz, Lebowitz, and Spohn~\cite{lebowitz1983}.
However, 
the conceptual foundations trace back to earlier efforts 
within the statistical mechanics community to construct consistent models 
of nonequilibrium systems 
interacting with multiple reservoirs~\cite{lebowitz1955,PhysRevE.76.031132,maes2021}.
The key insight is that when multiple reservoirs promote distinct transitions, 
local thermodynamic constraints naturally arise at the level of individual transitions.
These constraints depend on the specific quantities, such as energy, entropy, or matter, 
that are exchanged with the corresponding reservoir.
They provide a constructive principle 
for building physically sound models  of nonequilibrium behavior 
and 
ensure that detailed balance, i.e., the existence of an equilibrium state, 
is recovered in the absence of nonconservative forces.

Nowadays, 
local detailed balance lies at the core of nonequilibrium formulations of thermodynamics~\cite{crooks1998,lebowitz1999, seifert2012, maes2021, Rao2018a, falasco2021ldb, falasco2025RMP, Peliti2021}, including in chemical processes~\cite{Rao2016,Rao2018b,Avanzini2021}.


\section{Discussion and Further Perspectives\label{sec:conclusions}}

In this paper,
we established a general thermodynamic theory of
molecular systems undergoing thermally and photo-induced transitions.
By building on the last historical developments 
highlighted in Section~\ref{sec:micro_rev},
we showed that 
local detailed balance can be preserved even when combining 
thermally and photo-induced transitions, 
thus unifying two historically important forms of local detailed balance in the context of chemistry and radiation.

We further developed two sequential thermodynamically consistent coarse-graining procedures.
The first eliminates vibrational states
by assuming their rapid equilibration within electronic states.
The second eliminates electronic excited states
by assuming their rapid relaxation to a, in general, nonequilibration steady state.
The final result is a thermodynamic theory for photo-isomerization reactions
interconverting ground-state species,
which are proven to satisfy the local detailed balance conditions
just as the elementary transitions do.
This theory allowed us to shed light on the working mechanism 
of photo-driven ratchets.
Namely, we showed that
they operate exclusively through an information ratchet mechanism 
when powered by a constant light source 
in the same way 
that chemically driven ratchets do 
when powered by a constant chemical potential gradient.

Although we explicitly considered here only unimolecular transitions,
thermally induced reactions between ground state species of arbitrary molecularity
can be included in the theory too
(similarly to what has been done in Ref.~\cite{Penocchio2022}).
At the current stage, 
the theory cannot consider multimolecular transitions 
between excited- and ground-state species
like those involved in bimolecular quenchings, photo-additions, and photo-dissociations.

We furthermore focused on photo-driven systems,
where light is the sole source of free energy.
Nevertheless, 
our theory can be straightforwardly combined with 
thermodynamic theories for chemically driven systems~\cite{Gaspard2004, Schmiedl2007, Rao2018b, Qian2005, Rao2016, Avanzini2021, avanzini2022},
thus allowing the characterization of systems 
with both photo-driven and chemically driven reactions.
In such systems,
photons can be transduced into chemical energy 
and vice versa.
Analyzing these transduction processes in terms of thermodynamic gears~\cite{bilancioni2024elementaryfluxmodescrn, bilancioni2025energytransductioncomplexnetworks} 
may pave the way toward 
systematic studies of efficiency in photosynthetic systems.

Finally,
we stress that the working mechanism of photo-driven ratchets
could only be correctly identified by applying
the local detailed balance condition to
the exact expression of the ratcheting constant in Eq.~\eqref{eq:kr_def}.
Applying the same condition to
the approximate expression in Eq.~\eqref{eq:Kr_simplified}
fails to reveal the same working mechanism.
This discrepancy arises from neglecting certain reaction fluxes in Eq.~\eqref{eq:Kr_simplified}
that, although usually negligible in magnitude, are not independent of the remaining ones.
As a result, their omission obscures the constraints imposed by thermodynamics
on the ratcheting constant 
in the same way as overlooking 
local detailed balance conditions 
can lead to
incorrect conclusions about chemically driven systems~\cite{Blackmond2009, Astumian:2012aa}.


\section{Acknowledgments}

FA is supported by the project P-DiSC\#BIRD2023-UNIPD 
funded by the Department of Chemical Sciences of the University of Padova (Italy).
ME is supported by the Fond National de la Recherche-FNR, Luxembourg, 
by the Project ChemComplex (Grant No. C21/MS/16356329) 
and 
by project INTER/FNRS/20/15074473 funded by F.R.S.-FNRS (Belgium) and FNR.


\section*{Data Availability}

Data sharing is not applicable to this article as no new data were created or analyzed in this study.



\appendix


\section{Ratcheting Constant of Chemically Driven Ratchets\label{app:CDratchets}}

\begin{figure}
    \centering
    \includegraphics[width=0.49\textwidth]{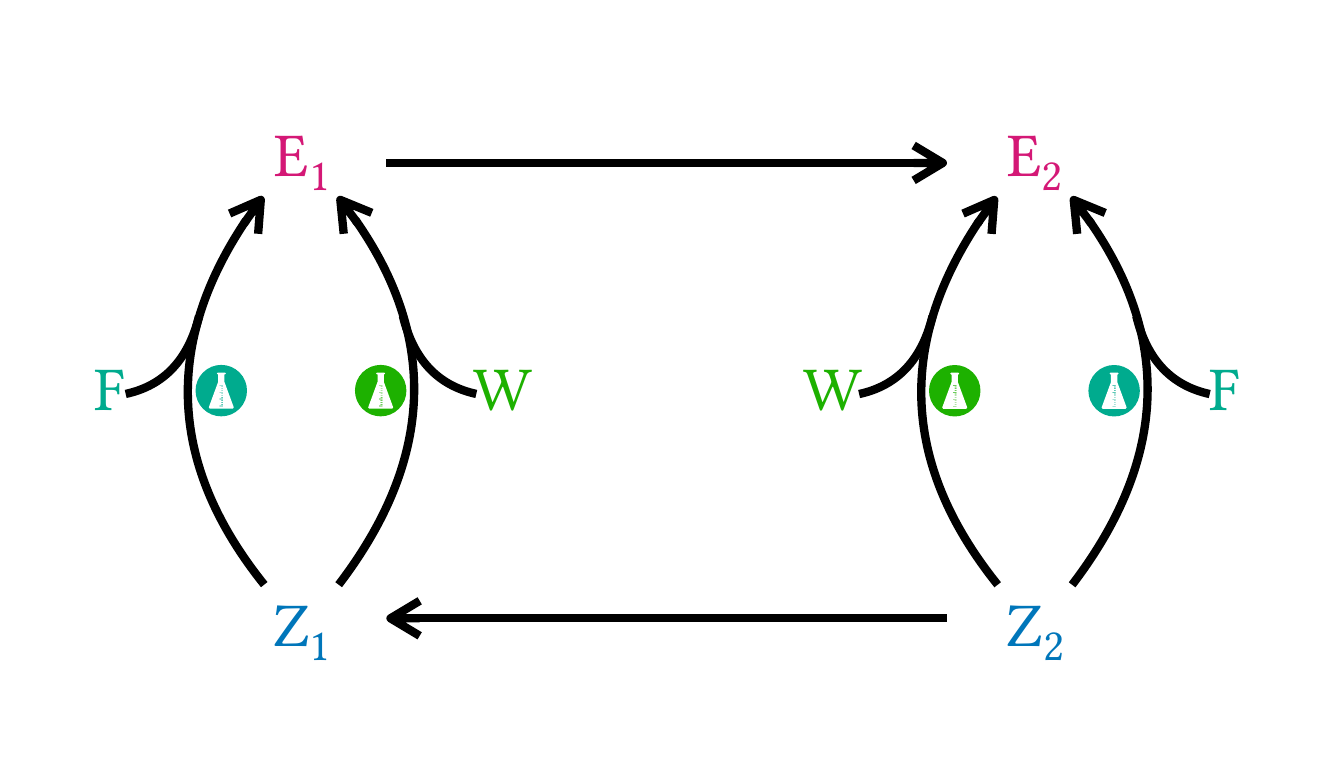}
    \caption{
    \Reaction\ mechanism of a prototypical chemically driven molecular \ratchet .
    All reactions are thermally induced and involve only ground-state species.
    The horizontal arrows represent the \reactions\ interconverting 
    $\cZa$ into $\cZb$, on the one hand, and $\cEa$ into $\cEb$, on the other hand.
    The other arrows (close to a flask) represent \reactions\ 
    coupled to chemostats (i.e., chemical reservoirs of either $\cF$ or $\cW$)
    interconverting $\cZa$ into $\cEa$, on the one hand, and $\cZb$ into $\cEb$, on the other hand.
    Each \reaction\ has always a backward counterpart, even if it is not represented.
    \label{fig:cdMM}}
\end{figure}

We consider a prototypical chemically driven molecular ratchet~\cite{ragazzon2018,Ragazzon2023review, Borsley2024,astumian2024review},
whose network of reactions is illustrated in Fig.~\ref{fig:cdMM},
and recap how the existence of a local detailed balance
condition 
affects 
its ratcheting constant $\Kr$.

We start by recalling that, according to mass-action kinetics,
the reaction fluxes read
\begin{subequations}
\begin{align}
\cdflux{\cZa}{\cEa}{\cF} &= \kcd{\cZa}{\cEa}{\cF}\conc{\cF} \conc{\cZa} \,,\\
\cdflux{\cEa}{\cZa}{\cF} &= \kcd{\cEa}{\cZa}{\cF} \conc{\cEa} \,\\
\cdflux{\cZa}{\cEa}{\cW} &= \kcd{\cZa}{\cEa}{\cW}\conc{\cW} \conc{\cZa} \,,\\
\cdflux{\cEa}{\cZa}{\cW} &= \kcd{\cEa}{\cZa}{\cW} \conc{\cEa} \,\\
\cdflux{\cEa}{\cEb}{} &= \kcd{\cEa}{\cEb}{} \conc{\cEa} \,,\\
\cdflux{\cEb}{\cEa}{} &= \kcd{\cEb}{\cEa}{} \conc{\cEb} \,,\\
\cdflux{\cZb}{\cEb}{\cF} &= \kcd{\cZb}{\cEb}{\cF}\conc{\cF} \conc{\cZb} \,,\\
\cdflux{\cEb}{\cZb}{\cF} &= \kcd{\cEb}{\cZb}{\cF} \conc{\cEb} \,\\
\cdflux{\cZb}{\cEb}{\cW} &= \kcd{\cZb}{\cEb}{\cW}\conc{\cW} \conc{\cZb} \,,\\
\cdflux{\cEb}{\cZb}{\cW} &= \kcd{\cEb}{\cZb}{\cW} \conc{\cEb} \,\\
\cdflux{\cZa}{\cZb}{} &= \kcd{\cZa}{\cZb}{} \conc{\cZa} \,,\\
\cdflux{\cZb}{\cZa}{} &= \kcd{\cZb}{\cZa}{} \conc{\cZb} \,,
\end{align}
\label{eq:flux_CD}
\end{subequations}
and must satisfy the following local detailed conditions
\begin{subequations}
\begin{align}
\kb T \ln\frac{\cdflux{\cZa}{\cEa}{\cF}}{\cdflux{\cEa}{\cZa}{\cF}} & = -(\cp{\cEa} - \cp{\cZa} - \cp{\cF}) \,,\\
\kb T \ln\frac{\cdflux{\cZa}{\cEa}{\cW}}{\cdflux{\cEa}{\cZa}{\cW} } & = -(\cp{\cEa} - \cp{\cZa} - \cp{\cW}) \,,\\
\kb T \ln\frac{\cdflux{\cEa}{\cEb}{}}{\cdflux{\cEb}{\cEa}{} } & = -(\cp{\cEb} - \cp{\cEa}) \,,\\
\kb T \ln\frac{\cdflux{\cZb}{\cEb}{\cF}}{\cdflux{\cEb}{\cZb}{\cF}} & = -(\cp{\cEb} - \cp{\cZb} - \cp{\cF}) \,,\\
\kb T \ln\frac{\cdflux{\cZb}{\cEb}{\cW}}{\cdflux{\cEb}{\cZb}{\cW}} & = -(\cp{\cEb} - \cp{\cZb} - \cp{\cW}) \,,\\
\kb T \ln\frac{\cdflux{\cZa}{\cZb}{}}{\cdflux{\cZb}{\cZa}{}} & = -(\cp{\cZb} - \cp{\cZa}) \,.
\end{align}
\label{eq:LDB_CD}%
\end{subequations}
Here, the species $\cF$ and $\cW$ are chemostats, 
namely, they represent the source of free energy the chemically driven ratchet uses to stay out of equilibrium.
Their concentrations, as well as their chemical potentials, are therefore treated as constant (control) parameters.
Furthermore, 
all reactions are thermally induced, but we dropped the superscript $\mathrm{q}$ 
to specify the chemostats they are coupled to ($\cF$ or $\cW$ or neither).

For the chemically driven molecular ratchet in Fig.~\ref{fig:cdMM},
the ratcheting constant reads
\begin{equation}
\begin{split}
\Kr = &\frac{
\bigg(
\cdflux{\cZa}{\cEa}{\cF} + \cdflux{\cZa}{\cEa}{\cW}
\bigg) 
\cdflux{\cEa}{\cEb}{} 
}{
\bigg(
\cdflux{\cEa}{\cZa}{\cF} + \cdflux{\cEa}{\cZa}{\cW} 
\bigg) 
\cdflux{\cEb}{\cEa}{}
}\times \\
&\frac{
\bigg(
\cdflux{\cEb}{\cZb}{\cF} + \cdflux{\cEb}{\cZb}{\cW} 
\bigg) 
\cdflux{\cZb}{\cZa}{} 
}{
\bigg(
\cdflux{\cZb}{\cEb}{\cF} + \cdflux{\cZb}{\cEb}{\cW}
\bigg) 
\cdflux{\cZa}{\cZb}{}
}
\,,
\end{split}
\label{eq:kr_def_CD}%
\end{equation}
and becomes 
\begin{equation}
\begin{split}
\Kr = \frac{
\bigg(
\frac{\cdflux{\cZa}{\cEa}{\cF}}
{\cdflux{\cZa}{\cEa}{\cW}} 
 + 1
\bigg) 
\bigg(
\frac{\cdflux{\cZb}{\cEb}{\cF}}
{\cdflux{\cZb}{\cEb}{\cW}} 
e^{\frac{\cp{\cW} - \cp{\cF}}{\kb T}}
 + 1
\bigg) 
}{
\bigg(
\frac{\cdflux{\cZa}{\cEa}{\cF}}
{\cdflux{\cZa}{\cEa}{\cW}} 
e^{\frac{\cp{\cW} - \cp{\cF}}{\kb T}}
 + 1\bigg) 
\bigg(
\frac{\cdflux{\cZb}{\cEb}{\cF}}
{\cdflux{\cZb}{\cEb}{\cW}} 
+ 1
\bigg) 
}
\,,
\end{split}
\label{eq:kr_def_CD_ldb}%
\end{equation}
by applying the local detailed balance conditions in Eq.~\eqref{eq:LDB_CD}.
Notice that $\Kr$ in Eqs.~\eqref{eq:kr_def_CD} and~\eqref{eq:kr_def_CD_ldb}
is independent of the concentrations $\conc{\cZa}$, $\conc{\cZb}$, $\conc{\cEa}$, $\conc{\cEb}$
even if it is written in terms of reaction fluxes rather than kinetic constants
(see Eq.~\eqref{eq:flux_CD}).
We chose to write $\Kr$ in terms of reaction fluxes
for consistency with what we have done in the main text 
for photo-driven ratchets in  Eqs.~\eqref{eq:kr_def} and~\eqref{eq:Kr_ldb}.

Finally, 
we recap the properties of the working mechanism of chemically driven ratchets that follow from Eq.~\eqref{eq:kr_def_CD_ldb}.
Crucially, these are the same properties we identified for photo-driven ratchets.
First, if $\cp{\cW} - \cp{\cF} = 0$, then $\Kr = 1$.
Namely, if the chemostats are in equilibrium, 
there are no (nonconservative) thermodynamic forces maintaining reactions out of equilibrium 
and directionality cannot emerge.
Second, 
$\Kr$ is independent of the reaction fluxes
$\cdflux{\cEa}{\cEb}{}$, $\cdflux{\cEb}{\cEa}{}$, $\cdflux{\cZb}{\cZa}{}$, and $\cdflux{\cZa}{\cZb}{}$
and, therefore, of the equilibrium constants 
\begin{subequations}
\begin{align}
\kqeq{\cEa}{\cEb}  & =\frac{\kcd{\cEa}{\cEb}{}}{\kcd{\cEb}{\cEa}{}}\,, \\
\kqeq{\cZa}{\cZb}  & =\frac{\kcd{\cZa}{\cZb}{}}{\kcd{\cZb}{\cZa}{}}\,.
\end{align}
\label{eq:Keq_rm_CD}
\end{subequations}
Namely, directionality in chemically driven \ratchets\
does not rely on the energy differences between reaction intermediates.
Third, if 
\begin{subequations}
\begin{align}
{\cdflux{\cZa}{\cEa}{\cF}}/\conc{\cZa} & = {\cdflux{\cZb}{\cEb}{\cF}}/\conc{\cZb} \,,\\
{\cdflux{\cZa}{\cEa}{\cW}}/\conc{\cZa} & = {\cdflux{\cZb}{\cEb}{\cW}}/\conc{\cZb}  \,,
\end{align}
\end{subequations}
then $\Kr = 1$.
Namely, directionality in chemically driven \ratchets\
relies on differences in the fluxes of the free-energy-harnessing reactions.

\newpage
\bibliography{biblio}
\end{document}